\documentclass[%
reprint,
superscriptaddress,
floatfix,
amsmath,
amssymb,
aps,
notitlepage
]{revtex4-2} 

\usepackage{graphicx}
\usepackage{bm}
\usepackage{xcolor}
\usepackage{natbib}
\usepackage{enumitem}
\usepackage{float}
\usepackage[colorlinks=true,linkcolor=blue,citecolor=blue,urlcolor=blue]{hyperref}

\graphicspath{{figures/}{auto/}}

\definecolor{greenish}{RGB}{108, 200, 105}
\definecolor{reddish}{RGB}{174,12,48}
\definecolor{blueish}{rgb}{0.12, 0.56, 1.0}
\definecolor{magentaish}{RGB}{242, 80, 93}
\hypersetup{
  colorlinks   = true,
  urlcolor     = greenish,
  linkcolor    = blueish,
  citecolor    = magentaish
}

\newcommand{\rmax}{\varrho_{\max}}          
\newcommand{\Qtot}{Q}          
\newcommand{\Qt}{\mathcal{Q}}    
\newcommand{\Cc}{\mathcal{C}}            
\newcommand{\ctot}{\Cc_{\rm tot}}          
\newcommand{\dd}{\mathrm{d}}
\newcommand{\Om}{\Omega}
\newcommand{\OF}{\Omega_F}
\newcommand{\ON}{\Omega_N}
\newcommand{\xt}{\tilde{x}}
\newcommand{\Xv}{\mathbf{X}}
\newcommand{\Ut}{\tilde{U}}
\newcommand{\asym}{\mathcal{A}}
\newcommand{\traf}{\mathcal{T}}        
\newcommand{\xv}{\mathbf{x}}

\begin{document}

\title{Emergent stigmergy and pheromone-guided decisions in a foraging collective}
\author{Manoj N H}
\affiliation{Department of Electrical Engineering, IIT Madras, Chennai TN 600036.}
\author{S Ganga Prasath}
\email{sgangaprasath@smail.iitm.ac.in}
\affiliation{Department of Applied Mechanics, IIT Madras, Chennai TN 600036.}

\date{}

\begin{abstract}
Ants in a colony forage and recruit other nest-mates for cooperative transport of food through stigmergy. Traditional approach to capture the foraging dynamics relies on fine-tuning phenomenological rules for pheromone deposition, chemotaxis, evaporation, and behavioral switching after reaching the food source. Here, we show that stigmergy emerges from an infinite-horizon stochastic optimal control problem involving outbound foragers traveling from the nest to the food source and returning agents traveling from the food source to the nest. Pheromone dynamics along a trail emerge from the forward propagating Hamilton--Jacobi--Bellman (HJB) equation, with the Bellman discount rate setting the rate of pheromone evaporation. Using this mapping, we identify the conditions under which trail feedback suppresses exploration or leads to the exploitation of established trails. For ants exploring multiple trails, we also find flux-potential relations, analogous to Ohm's law, with ant flux and inverse of squared value field playing the roles of current and resistance on the network. Our model captures collective behavior observed in the classic double-bridge experiments: trail selection on identical routes through a symmetry-breaking instability, selection of the shorter bridge, and bistability in the trail-length ratio that can leave ants transiently committed to the longer route. Our formulation places collective foraging in the broader framework of stochastic optimal control, while simultaneously providing an unifying picture on their emergent behavior consistent with experimental observations.
\end{abstract}

\maketitle

\noindent\textit{Introduction}---Ant colonies discover and exploit foraging trails even though individuals hold only local information. They achieve this via \textit{stigmergy}~\cite{Grasse1959, Theraulaz1999, Heylighen2016}, the process by which successful foragers lay pheromone to recruit nestmates through chemo-sensing. The positive feedback set by returning ants depositing a volatile pheromone biases the motion of the outbound ants up the gradient, while the evaporation discards stale information~\cite{Dorigo2004,Boissard2013,Amorim2015,Perna2012}. The collective leverages the type of pheromone and the deposition rate to select a worthwhile trail to successfully transport food to the nest. Classical controlled experiments~\cite{Hangartner1969, Goss1989, Sumpter2003, Deneubourg1990} on bridges of different lengths between food and nest (which force the collective to choose a trail) to understand the decision making process point to: (a) no preference in the trail but commitment to a single trail even when the trails are of the same length, (b) preference towards the shortest trail, given an option between shorter and longer one, (c) trapping in the long trail when a shorter route is made available after commitment.

Although phenomenological approaches remain the dominant paradigm for modeling the dynamics of such search and exploitation problems, navigation strategies have also been derived as the solution of a search game~\cite{krishnan2026stigmergic, Vergassola2007, heinonen2023optimal, Pezzotta2018}, and recently reinforcement-learning formulations of collective chemotaxis have reproduced Keller--Segel dynamics~\cite{Kato2025,KellerSegel,prasath2022dynamics}. The technical device in such treatments involves the Cole--Hopf transformation, which linearizes the Hamilton--Jacobi-Bellman (HJB) equation of a quadratically regularized control problem~\cite{Kappen2005,Todorov2009,Fleming2006}. In this work, we pose the collective foraging task as an infinite-horizon optimal control problem, where the foragers rely on information carried by returners with a reward for reaching the food source and an attractive homing potential that guides them towards the nest~\cite{Wehner2003, Wehner2020}. Our model on single trail shows: (i) emergence of stigmergy with the chemotaxis of foragers (Cole-Hopf transformed value field, sometimes referred as the \textit{desirability}, acting as the pheromone) and pheromone laying behavior by returners, (ii) transition from camping phase (with ants confined to the nest) to foraging phase (successfully reaching food source) with the phase boundary determined by the quality of food source and the colony parameters, (iii) an effective Ohm's law for the foragers and returners with the current and resistance determined by the ant density and the value field along the trail. In the double-bridge setting with trails marked by both foragers and returners, our model captures: (i) a symmetry breaking instability in identical trails in which ants commit to a particular trail -- driven by negative differential resistance, (ii) selection of the shorter trail over longer one -- owing to lower resistance, (iii) bistability in the commitment when the trail length ratio is not unity -- consistent with experimental observations~\cite{Deneubourg1990, Hangartner1969, Goss1989}.\\

\begin{figure*}[t]
\centering
\includegraphics[width=\textwidth]{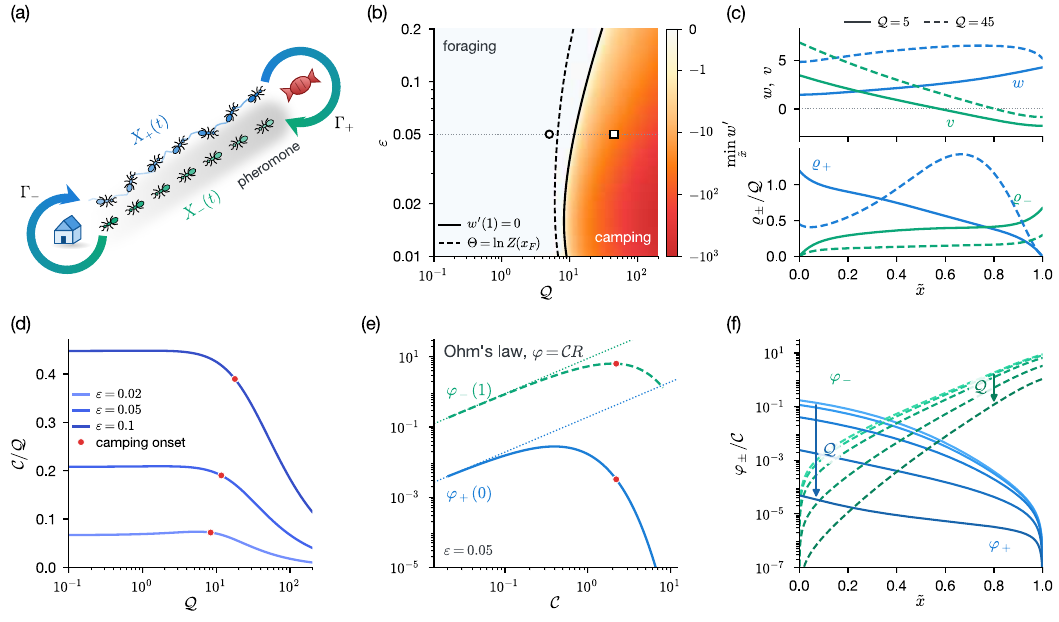}
\caption{\textbf{Single trail foraging and emergence of Ohm's law.} (a)~Foraging ants $X_+(t)$ walk from the nest to the food, where they switch to returners $X_-(t)$ at rate $\Gamma_+$; returners switch back at the nest at rate $\Gamma_-$ and lay the trail that onward ants follow. (b)~Phase diagram of foraging on a single trail: Contour plot showing the smallest outbound drift $\min_{\xt}w'$ in the plane of colony size $\Qt$ and scaled diffusion $\varepsilon$. The smallest drift vanishes in the foraging phase, where $w$ rises monotonically to the food, and is negative once $w$ develops an interior maximum (camping). Solid line: camping onset $w'(1)=0$; dashed line: equality in the necessary condition Eq.~\eqref{eq:theta}, $\Theta=\ln Z(x_F)$; markers: the two steady-states shown in~(c). See SM Video 1 for the dynamics in foraging and camping phase. (c)~Value fields $w=\ln Z$ and $v=\ln\Psi$ (top) and densities per ant $\varrho_\pm/\Qt$ (bottom) at $\varepsilon=0.05$, in the foraging ($\Qt=5$, solid) and camping ($\Qt=45$, dashed) phases. The maximum density of returners is at the boundary in the foraging phase and in the bulk in the camping phase. (d)~Circulation per ant $\Cc/\Qt$ against $\Qt$ for three values of $\varepsilon$; dots mark the camping onset. (e)~Ohm's law: nest potential $\varphi_+(0)=\Cc\,R^+$ and food potential $\varphi_-(1)=\Cc\,R^-$, Eqs.~\eqref{eq:ohmlocal}--\eqref{eq:rhoclosed}, against the circulation $\Cc$ as $\Qt$ is swept at $\varepsilon=0.05$. Dotted lines: Ohmic response with the zero-current resistances; dots: camping onset. Both potentials pass through a maximum, beyond which the trail carries more current at a lower potential. (f)~Potential profiles $\varphi_+/\Cc=\int_{\xt}^{1}e^{-2w}$ (solid) and $\varphi_-/\Cc=\int_0^{\xt}e^{-2v}$ (dashed) at $\varepsilon=0.05$ for $\Qt=0.3,1,3,10,30$; arrows point towards increasing $\Qt$. All the results are at $g_F=6$, $g_N=2$ and $\tilde\kappa=4$ (see SM Secs.~\ref{sm:conditions},~\ref{sm:onedim} for details).}%
\label{fig:single}
\end{figure*}

\noindent\textit{Model}---Ants move in $\Om\subseteq\mathbb R^2$, which contains a nest region $\ON$ and a food region $\OF$ (see Fig.~\ref{fig:single}(a) for a schematic). Each ant stays in a behavioral state $S_t\in\{+,-\}$, where $+$ denotes an outbound foraging ant traveling from nest towards food and $-$ a returning one, so that the full state is the Markov process $\mathbf{Y}_t=(\Xv_t,S_t)$ with
\begin{equation}
\dd \Xv_t=\bm u_\pm(\Xv_t,t)\,\dd t+\sqrt{2D}\,\dd \bm W_t .
\label{eq:sde}
\end{equation}
Ants can switch between states through position-dependent jump process, $\mathbb P(S_{t+\dd t}=\mp\,|\,S_t=\pm,\Xv_t=\xv)=\Gamma_\pm(\xv)\,\dd t$, with $\Gamma_+$ defined near $\OF$ and $\Gamma_-$ near $\ON$ (shown schematically in Fig.~\ref{fig:single}(a)). We simplify the model by assuming the diffusivity $D$ and the mobility $\beta$ to be the same in both states. Here $\beta$ sets the cost of control through the running cost $|\bm u|^2/2\beta$, so large $\beta$ means cheap control.

Each agent minimises a discounted cost up to its next switching time, taking the opposite value function as the terminal payoff. The value field $V^\pm(\xv,t)$ is the minimum expected cost for an ant in state $\pm$ at position $\xv$ and time $t$, and it obeys the coupled Bellman equations,
\begin{align}
V^{+}&=\min_{\bm u_+}\mathbb E\bigg[\!\int_t^{\tau_+}\!\!\!\dd s\,e^{-\lambda (s-t)}
\bigg(\dfrac{|\bm u_+|^2}{2\beta}-\alpha\varrho_-\bigg)
\nonumber\\[-2pt]
&\hspace{2.1cm}+e^{-\lambda(\tau_+-t)}\big(V^--r_F\big)\bigg],
\label{eq:Vp}
\end{align}
\begin{align}
V^{-}&=\min_{\bm u_-}\mathbb E\bigg[\!\int_t^{\tau_-}\!\!\!\dd s\,e^{-\lambda (s-t)}
\bigg(\dfrac{|\bm u_-|^2}{2\beta}+U\bigg)
\nonumber\\[-2pt]
&\hspace{2.1cm}+e^{-\lambda(\tau_--t)}\big(V^+-r_N\big)\bigg].
\label{eq:Vm}
\end{align}
Here $\varrho_-$ is the density of ants returning to the nest from the food source, and the value field for the foragers consists of a positive running reward at rate $\alpha\varrho_-$ for being where the returners are and also collecting $r_F$ when reaching food. We assume a simple attractive homing potential $U(\xv) = \kappa |\xv|^2/2$, with $\kappa$ being the stiffness of the potential, that guide the returners towards the nest and they also collect reward $r_N$ at the nest. It is important to note that no pheromone appears in the specification of the problem and the reward structure is one-way, in that returners are followed but outbound ants are not.\\

\noindent \textit{Coupled HJB equations and the sensing law}---Applying Bellman's principle across the switching event (see SM Sec.~\ref{sm:hjb}) gives the coupled system
\begin{align}
\partial_t V^{\pm}+D\nabla^2V^{\pm}-\frac{\beta}{2}|\nabla V^{\pm}|^2
-\lambda V^{\pm}+f_\pm
\nonumber\\
+\,\Gamma_\pm\big(V^{\mp}-V^{\pm}-\delta_\pm\big)=0,
\label{eq:hjb}
\end{align}
with running costs $f_+=-\alpha\varrho_-$ and $f_-=U$, conversion rewards $\delta_+=r_F$ and $\delta_-=r_N$, and optimal feedback $\bm u^*_\pm=-\beta\nabla V^\pm$. The Cole--Hopf map
\begin{equation}
V^+=-\frac{2D}{\beta}\ln Z,\qquad V^-=-\frac{2D}{\beta}\ln\Psi,
\label{eq:coleHopf}
\end{equation}
cancels the quadratic gradient term identically and turns the feedback into
\begin{equation}
\bm u^*_+=2D\,\nabla\ln Z,\qquad \bm u^*_-=2D\,\nabla\ln\Psi .
\label{eq:logsensing}
\end{equation}
The optimal action an agent takes is to climb the logarithm of the field with drift velocity $2D\nabla Z/Z$ (which is nothing but Weber--Fechner chemosensing~\cite{Kalinin2009,Lazova2011}).

Substituting Eq.~\eqref{eq:coleHopf} into Eq.~\eqref{eq:hjb} (derived in SM Sec.~\ref{sm:colehopf}) gives the exact transformed dynamics for the value fields that guide the agents towards the food and the nest as
\begin{align}
\partial_t Z&=-\Big(D\nabla^2Z+\frac{\alpha\beta}{2D}\varrho_-Z-\lambda Z\ln Z\Big)
\nonumber\\
&\qquad-\Gamma_+Z\Big(\!-\!\ln Z+\ln\Psi+\frac{\beta r_F}{2D}\Big),
\label{eq:Zfull}\\
\partial_t \Psi&=-\Big(D\nabla^2\Psi-\frac{\beta}{2D}U\Psi-\lambda \Psi\ln \Psi\Big)
\nonumber\\
&\qquad-\Gamma_-\Psi\Big(\!-\!\ln \Psi+\ln Z+\frac{\beta r_N}{2D}\Big).
\label{eq:Psifull}
\end{align}

\noindent\textit{Collective dynamics}---The stochastic dynamics of foragers and returners in Eq.~\eqref{eq:sde} can be converted to collective dynamics in the form of a density evolution equation (derived in SM Sec.~\ref{sm:fpe})
\begin{align}
\partial_t\varrho_\pm + 2D\nabla\!\cdot\!\big(\varrho_\pm\nabla\ln \chi_\pm\big) =
D\nabla^2\varrho_\pm\mp\Gamma_+\!\varrho_+\!\pm\!\Gamma_-\varrho_-,
\label{eq:fpe}
\end{align}
with $\chi_+=Z$, $\chi_-=\Psi$. Eqs.~\eqref{eq:Zfull}--\eqref{eq:fpe} form a closed forward--backward system: $\varrho_-$ sources $Z$, and $\nabla\ln Z$ transports $\varrho_+$.

We see that wherever the outbound current $\bm J_+=2D\varrho_+\nabla\ln Z-D\nabla\varrho_+$ vanishes, integration gives
\begin{equation}
\varrho_+\;\propto\;Z^{2}\;=\;e^{-\beta V^+/D},
\label{eq:boltzmann}
\end{equation}
which is a Boltzmann distribution at temperature $D/\beta$ over the value landscape. In quiescent parts of the domain, or in a colony at rest, the standing density of outbound ants therefore scales as the square of the pheromone field. In a circulating colony, Eq.~\eqref{eq:boltzmann} is not the state but the reference state, and the departure from it is what carries the traffic and is the content of the circuit reduction shown later.\\

\noindent\textit{Value field acts as pheromone}---Away from the switching regions, Eq.~\eqref{eq:Zfull} with forward in time propagation reads
\begin{equation}
\partial_t Z=D\nabla^2Z+\underbrace{\frac{\alpha\beta}{2D}\varrho_-Z}_{\text{deposition}}
-\underbrace{\lambda Z\ln Z}_{\text{degradation}} .
\label{eq:pheromone_exact}
\end{equation}
All the terms are local while the source is proportional to the density of returning ants, and the sink is the Bellman discount. Interestingly, discounting appears as evaporation because the two have the same significance: a reward $R$ realized at time $t$ enters Bellman's objective weighted by $e^{-\lambda t}$, while a quantity $q$ obeying $\dot q=-\lambda q$ with $q(0)=R$ has $q(t)=Re^{-\lambda t}$. Discounting a stored reward is therefore indistinguishable from letting an undiscounted store decay exponentially, and a population of molecules each degrading at a constant rate $\lambda$ realizes exactly this.

The HJB equation in Eq.~\eqref{eq:hjb} runs backward in time: the optimal value at time $t$ is set by costs that occur later, so solving it requires knowing the future which neither an ant can perform nor does a chemo-attractant obey. A deposited field can only evolve forward, accumulating what has already been laid by active agents. We therefore propagate Eq.~\eqref{eq:pheromone_exact} forward in time and at stationarity this changes nothing, because the time-independent solutions of the forward and backward equations are identical. The forward equation for $Z$, however, relaxes to the optimal value on a time $\sim\lambda^{-1}$, which is the pheromone lifetime and the colony does not solve the problem in advance; it computes the solution iteratively, using the substrate as its memory. The forward field integrates the past deposition instead of anticipating future returns. The resulting policy is therefore optimal only once the trail has relaxed, and it lags any change in the environment by roughly one pheromone lifetime. Such an assumption is valid in the foraging task as trail selection and transport along a trails lasts much longer than the time taken to establish the trail.

On the other hand, linearizing Eq.~\eqref{eq:pheromone_exact} by setting $\phi=Z-1$ and expanding recovers the canonical pheromone equation (details in SM Sec.~\ref{sm:weaktrail}) $\partial_t\phi=D\nabla^2\phi+ ({\alpha\beta}/{2D})\varrho_--\lambda\phi$. But the exact equation shows that deposition is autocatalytic, at a rate proportional to the trail already present, and that degradation is superlinear, with an effective evaporation rate $\lambda\ln Z$ that grows with concentration (under the condition that $Z \geq 1$, shown in SM Sec.~\ref{sm:positivity}), which we show below is physically relevant and is absent in the linearized version.\\

\noindent\textit{Dynamics on a single trail}---We now solve the one-dimensional problem on a segment of length $L$, with the nest at $x_N \equiv x=0$ and the food at $x_F \equiv x=L$, each co-located with a reflecting wall. The independent non-dimensional groups are scaled diffusion $\varepsilon=D/\lambda L^2$, the two conversion rewards $g_{F,N}=\beta r_{F,N}/2D$, the scaled colony size $\Qt=\alpha\beta\Qtot/(2D\lambda L)$, and the strength of the homing potential $\tilde\kappa=\beta\kappa L^{2}/4D\lambda$. In the logarithmic variables $w=\ln Z$ and $v=\ln\Psi$ the steady bulk equations~\ref{eq:Zfull} and~\ref{eq:Psifull} are nothing more than the HJB equations and now read $\varepsilon(w''+w'^2)-w+\varrho_-=0$ and $\varepsilon(v''+v'^2)-v-\Ut=0$ (see SM Sect.~\ref{sm:onedim} for details). From hereon $\varrho_\pm$ and $J_\pm$ are scaled, but we keep the dimensional symbols. The outbound field $w$ increases towards the food and the returner field $v$ decreases away from the nest; the densities satisfy $\varrho_+(1)=\varrho_-(0)=0$; and the optimal drifts $w'$ and $v'$ are bounded and of opposite sign, as they must be for a circulating colony. The reward $\alpha\varrho_-$ in Eq.~\eqref{eq:Vp} is a shaping term that encourages trail following behavior but it is not the goal, and if it is too strong, the optimal policy is to remain at the pheromone maximum indefinitely. For foragers to successfully reach the food source, the cost of camping must be higher than that of reaching the food source. We can bound the cost of any food-reaching strategy from below by the cost of camping at $\rmax=\max_x \varrho_-$ and it gives the necessary condition (derived in SM Sec.~\ref{sm:conditions})
\begin{equation}
\Theta\;\equiv\;\frac{\alpha\beta\rmax}{2D\lambda}\;<\;\ln Z(x_F).
\label{eq:theta}
\end{equation}
Thus successful foraging requires $Z\gg1$ whenever trails are appreciable and the linearization $\phi=Z-1\ll1$ that recovers the canonical pheromone equation requires $Z\to1$ --- the linear pheromone equation is valid only far from the food-source or in the regime where foraging fails. In the foraging phase, given $Z\ge1$ (detailed in SM Sec.~\ref{sm:positivity}), this bounds $Z^{-2}\le1$, and as we shall see, limits the \textit{resistance} of the trail forcing \textit{zero-current} region to carry no traffic. It is worth noting that $\Theta$ can only be estimated after solving for the densities as $\rmax$ can be determined from the steady-state solution.

The collective dynamics along the trail can be characterized via a phase-diagram with scaled diffusion $\varepsilon$ and colony size $\Qt$ as the axes (shown in Fig.~\ref{fig:single}(b), see also SM Fig.~\ref{fig:phase}). Successful foraging corresponds to density (and value field) peaks at the food source and nest of returners and foragers with continuous circulation along the trail, while camping/aggregation results in a maximum in the bulk with reduction in current (shown in Fig.~\ref{fig:single}(c, d), see SM Video 1 for the dynamics). The transition from the foraging phase to the camping phase is captured by the outbound drift $w'(x)$ at the food location, which is $w'(1) \geq 0$ for the foraging phase and becomes $w'(1) < 0$ for the camping phase (shown in Fig.~\ref{fig:single}(b) and SM Fig.~\ref{fig:phase}). Equation~\eqref{eq:theta} acts as a lower bound on the phase boundary since this estimate does not take into account the diffusive effects (discussed in SM Sec.~\ref{sm:conditions}).\\

\begin{figure*}[t]
\centering
\includegraphics[width=\textwidth]{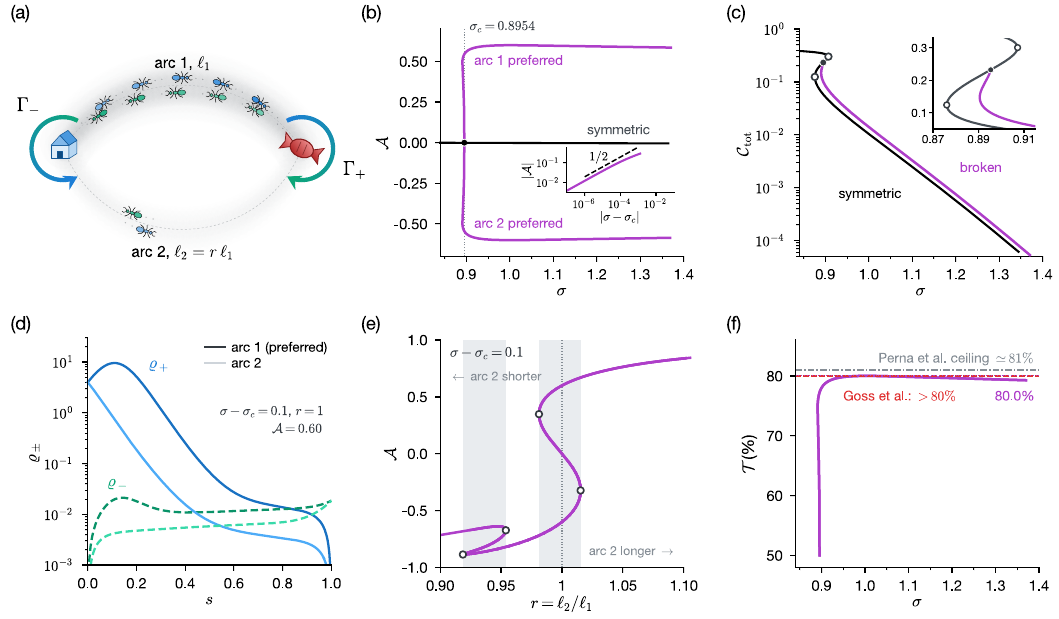}
\caption{\textbf{Trail selection on the double bridge with bidirectional marking.} (a)~Two trails of lengths $\ell_1$ and $\ell_2=r\ell_1$ join the nest to the food. Returners mark at unit rate and onward ants at rate $\sigma$, so both read the mark density $m=\varrho_-+\sigma\varrho_+$. (b)~Traffic asymmetry $\asym$ from Eq.~\eqref{eq:asym}, against $\sigma$ on a bridge of equal length: the symmetric branch and the pair of broken branches meet at the pitchfork $\sigma_c=0.8954$ (filled dot). The pitchfork is weakly subcritical with the symmetric branch being unstable beyond $\sigma_c$: each broken branch turns at a fold at $\sigma=0.8905$, $|\asym|=0.48$, before growing to its maximum $|\asym|=0.60$. Inset: $|\asym|\sim|\sigma-\sigma_c|^{1/2}$ near the branch point. (c)~Total circulation $\ctot$ on the same branches fall exponentially with $\sigma$ at larger $\sigma$; open circles: folds of the symmetric branch at $\sigma=0.876$ and $0.907$, filled: the branch point. Inset: the fold region on linear axes. (d)~Densities $\varrho_+$ (solid) and $\varrho_-$ (dashed) along the preferred arc (dark) and the other arc (light), on the broken branch at $(\sigma-\sigma_c)=0.1$ and $r=1$, where $\asym=0.60$. (e)~Unfolding in the length ratio $r$ at $(\sigma-\sigma_c)=0.1$. Circles: folds; shaded: the ranges $0.919<r<0.954$ and $0.981<r<1.015$ in which several steady states coexist. (f)~Traffic fraction on the preferred arc $\traf$ reaches $80.0\%$, against the $>80\%$ measured by Goss \emph{et al.}~\cite{Goss1989} and the $\simeq81\%$ ceiling of the collective model of Perna \emph{et al.}~\cite{Perna2012}. Simulation results are at $\varepsilon=0.05$, $g_F=6$, $g_N=2$, $\tilde\kappa=4$, $\Qt=2.154$ and $\sigma_-=1$, using pseudo-arclength continuation~\cite{Auto07p}.}%
\label{fig:bridge}
\end{figure*}

\noindent\textit{Steady traffic obeys Ohm's law}---Away from the conversion regions neither species has a source, so in steady state each current is constant along a trail: $J_+=\Cc$ and $J_-=-\Cc$, with non-dimensional circulation $\Cc=\alpha\beta\,c/2D\lambda^{2}L$. Defining the potentials
\begin{equation}
\varphi_+\;\equiv\;\varepsilon\,\frac{\varrho_+}{Z^{2}},\qquad
\varphi_-\;\equiv\;\varepsilon\,\frac{\varrho_-}{\Psi^{2}},
\label{eq:potential}
\end{equation}
gives $\partial_{\xt}\varphi_+=Z^{-2}\big(\varepsilon\varrho_+'-2\varepsilon\varrho_+(\ln Z)'\big)=-Z^{-2}J_+$, so the transport equation for the foragers collapses to
\begin{equation}
-\,\partial_{\xt}\varphi_+\;=\;\mathcal{C}\,Z^{-2},
\label{eq:ohmlocal}
\end{equation}
and similarly $\partial_{\xt}\varphi_-=\mathcal{C}\,\Psi^{-2}$ for the returners. This is Ohm's law in differential form (see Fig.~\ref{fig:single}(e,f)): $\varphi_+$ is a potential that drops in the direction of foraging, $\mathcal{C}$ is the current, and $Z^{-2}$ is the local resistivity. Since $\varrho_+\propto Z^2$ is the zero-current Gibbs state, $\varphi_+$ measures the excess of the density over its local equilibrium weight, and it is this excess that drives traffic. Wherever the solution is locally Gibbs, $\varphi_+$ is constant and nothing flows.

Integrating Eq.~\eqref{eq:ohmlocal} with the ground $\varphi_+(1)=0$, set by $\varrho_+(1)=0$, gives the densities in closed form,
\begin{equation}
\varrho_+(\xt)=\frac{\mathcal{C}}{\varepsilon}\,Z^2(\xt)\!\int_{\xt}^1\!\frac{\dd y}{Z^2(y)},
\quad
\varrho_-(\xt)=\frac{\mathcal{C}}{\varepsilon}\,\Psi^2(\xt)\!\int_0^{\xt}\!\frac{\dd y}{\Psi^2(y)} .
\label{eq:rhoclosed}
\end{equation}
Figure~\ref{fig:single}(e) shows the nest and food potentials $\varphi_+(0)$ and $\varphi_-(1)$ against the circulation $\mathcal{C}$ as the colony size $\Qt$ changes. In the foraging phase the potentials obey Ohm's law, $\varphi_+(0)=\mathcal{C}R^+$ and $\varphi_-(1)=\mathcal{C}R^-$, with resistances $R^+=\int_0^1\!Z^{-2}\,\dd \xt$ and $R^-=\int_0^1\!\Psi^{-2}\,\dd \xt$. As $\Qt\to0$, $\varphi_+(0)$ and $\mathcal{C}$ both vanish while the resistances tend to constant zero-current values $R_0^{\pm}$, obtained by solving the zero-density problem (dotted lines in Fig.~\ref{fig:single}(e)). Both potentials pass through a maximum as $\mathcal{C}$ grows, $\varphi_+(0)$ already in the foraging phase and $\varphi_-(1)$ near camping onset, beyond which the trail carries more current at a lower potential. The potentials are not monotone because the resistivity $Z^{-2}\sim e^{-2w}$ is not constant -- it is exponentially small where the trail is strong, so a well-marked trail is an exponentially good conductor. This amplification replaces the power-law choice functions of phenomenological treatments~\cite{Goss1989,Deneubourg1990} and, as we shall see, plays a central role in symmetry breaking with bidirectional marking.\\

The Ohm's law relations for the foragers and returners can be extended to a network of trails. Let trail $i$ join a nest node to a food node and carry outbound and returning currents $\Cc_i^+$ and $\Cc_i^-$. Continuity of $\varrho_+$ at the nest removes the trail dependence from $\varphi_+(0)=\Cc_i^+R^+_i$, which gives the splitting law
\begin{equation}
\Cc_i^{+}=\ctot\,\frac{(R^+_i)^{-1}}{\sum_j (R^+_j)^{-1}},\qquad
\Cc_i^{-}=\ctot\,\frac{(R^-_i)^{-1}}{\sum_j (R^-_j)^{-1}},
\label{eq:ohm}
\end{equation}
with trail resistances
\begin{equation}
R^+_i=\int_0^{L_i}\!e^{-2w_i}\,\dd \xt,\qquad
R^-_i=\int_0^{L_i}\!e^{-2v_i}\,\dd \xt
\label{eq:resistances}
\end{equation}
and $\ctot=\sum_i \Cc_i^+=\sum_i \Cc_i^-$ the total circulation. Two consequences follow. First, splitting a trail at an interior point shows that resistances in series add, and Eq.~\eqref{eq:ohm} shows that they add reciprocally in parallel, so an arbitrary network reduces to circuit algebra once $w$ and $v$ are known, with no further partial differential equations to solve. All the nonlinearity sits in the dependence of an arc's resistivity on the current that arc carries. Second, outbound and returning traffic split independently i.e. an ant that reaches the food along one arc is under no obligation to return along it, and assuming otherwise over-determines the network: with a single current per arc, continuity of $\varrho_+$ at the nest gives $\Cc_1R^+_1=\Cc_2R^+_2$ and continuity of $\varrho_-$ at the food gives $\Cc_1R^-_1=\Cc_2R^-_2$, which are two conditions on one ratio and are compatible only in the non-generic case $R^+_1/R^-_1=R^+_2/R^-_2$.\\

Further, the circuit reduction of the model helps show that trail selection cannot happen when the onward foragers do not deposit pheromones. Let us consider the double-bridge problem (shown schematically in Fig.~\ref{fig:bridge}(a)), the variable dependencies on each trail follows,
\begin{equation}
\tilde{U}_i\rightarrow v_i\rightarrow R^-_i\rightarrow \Cc_i^{-}\rightarrow
\varrho_{-,i}\rightarrow w_i\rightarrow R^+_i\rightarrow \Cc_i^{+},
\label{eq:openchain}
\end{equation}
and there the chain stops. The returner network sets the resistivity seen by outbound ants, whereas the outbound network does not affect the returning ants, because only returners deposit and the outbound ants only read. The returner resistances $R^-_i$ are therefore fixed by the homing potential and the arc lengths alone, independently of the circulation, of the colony size. On a symmetric bridge, the $v$ problem has trail-exchange-symmetry, so we have $v_1\equiv v_2,\ \implies R^-_1=R^-_2$, hence $\varrho_{-,1}\equiv\varrho_{-,2},\ \implies w_1\equiv w_2, R^+_1=R^+_2$. The $50/50$ split is then the steady state at every colony size and no selection is possible. On an asymmetric bridge the split $\Cc_i^-\propto (R^-_i)^{-1}$, the shortest trail is selected because of least resistance once and for all by geometry, so there exists no bistability at any trail-length ratios.\\

\noindent\textit{Bidirectional marking closes the loop}--- Several ant species including \emph{L.\ humile}, mark with pheromones both when leaving and when returning to the nest, during foraging and exploratory recruitment~\cite{Deneubourg1990, Hangartner1969, Goss1989}. So each ant contributes and responds to the same chemical field. Let returners deposit at unit rate and outbound ants at relative rate $\sigma\ge0$, such that the mark density is $m=\varrho_-+\sigma\varrho_+$, and let returners be rewarded for marks with weight $\alpha_-$ alongside the outbound weight $\alpha$. The running costs become $f_+=-\alpha m$ and $f_-=U-\alpha_-m$, and the same Cole--Hopf step gives the steady instantaneous-switching equations in dimensionless logarithmic variables. We then have (see SM Sec.~\ref{sm:bidirectional} for details)
\begin{equation}
\begin{aligned}
\varepsilon\big(w''+w'^{2}\big)-w+\varrho_-+\sigma\varrho_+&=0,\\
\varepsilon\big(v''+v'^{2}\big)-v-\Ut+\sigma_-\varrho_-+\eta\varrho_+&=0,
\end{aligned}
\label{eq:bidirfields}
\end{equation}
with $\sigma_-=\alpha_-/\alpha$ the sensing asymmetry and $\eta=\sigma\sigma_-$ the cross-coupling. The four coefficients $(1,\sigma;\sigma_-,\eta)$ form a deposition--sensing matrix whose rows index who reads and whose columns index who deposits. Two parameters are added, both of order unity and setting either $\sigma$ or $\sigma_-$ to zero re-opens the chain Eq.~\eqref{eq:openchain}. With the bidirectional coupling, only the field equations change whereas the transport equations and the entire circuit reduction, Eqs.~\eqref{eq:potential}--\eqref{eq:resistances}, are unaffected.\\

\noindent\textit{Symmetry breaking and trapping on the double bridge}---We solve Eq.~\eqref{eq:bidirfields} on a two-arc graph mimicking the double bridge experiments (schematic in Fig.~\ref{fig:bridge}(a)) using pseudo-arclength continuation~\cite{Auto07p} and define the traffic asymmetry (see SM Sec.~\ref{sm:auto}),
\begin{equation}
\asym=\frac{\Cc_1^{+}-\Cc_2^{+}}{\Cc_1^{+}+\Cc_2^{+}},
\label{eq:asym}
\end{equation}
on the bridge as the order parameter. The fraction of outbound traffic on the preferred trail is, $\traf\equiv{\max_i (\Cc^{+}_i)}/{\ctot}={(1+|\asym|)}/{2},$ measures the strength of the selection. At small $\sigma$, outbound ants choose a trail by following marks that are laid almost entirely by returners. Any fluctuation in the marking leaves almost nothing behind, so the trail's resistance barely changes, and the next outbound ants see two nearly identical trails  (shown in SM Video 2). As we increase $\sigma$, those extra outbound ants now mark the chosen trail themselves and their choice becomes slightly better marked, which recruits the next ants, which mark it further. Choosing and marking are now done by the same ants, so the choice can feed on itself and is reflected in the form of a negative change in the differential potential $\delta \varphi_+(0)$ with an increase in the density of the agents or the circulation $\delta \Cc$, i.e. negative differential resistance. Given that the total number of ants is conserved, even on a symmetric bridge $r=1$, due to the exponential change in resistance, the system undergoes a pitchfork bifurcation at $\sigma_c=0.8954$ at at $\dd \varphi_+(0)/\dd \Cc=0$ with traffic asymmetry $|\asym|\sim|\sigma-\sigma_c|^{1/2}$ close to the branch point (see Fig.~\ref{fig:bridge}(b) and inset, also SM Video 2 for the dynamics). The pitchfork is weakly subcritical and each broken branch first bends back to a fold at $\sigma=0.8905$, and the outbound traffic on the preferred arc saturates at $\traf=80\%$ (shown in Fig.~\ref{fig:bridge}(d,f), as measured by Goss \emph{et al.}~\cite{Goss1989}). The saturation level rises with colony size, and the pitchfork turns subcritical just above $\Qt=2$. Such a collective commitment to a particular arc reproduces the observation that a bridge of equal arms is resolved strongly in each trial but with no preference across trials~\cite{Goss1989}. In the weak diffusion limit, $w(0) \approx w_0 + \sigma \varrho_+(0)$ and Ohm's law gives $\varrho_+(0) \approx (\Cc \ell/\varepsilon) e^{2(w(0) - w_{\min})}$ and in the strong trail regime, $\Qt = \varrho_+(0) \ell_N$ where $\ell_N$ is the width of the pile near the nest. We can then see that total circulation drops with an increase in deposition rate such as $\ctot \approx \Cc_* e^{-2 \Qt \sigma/\ell_N}$, as observed in Fig.~\ref{fig:bridge}(c).

When the arcs differ in length, $r=\ell_2/\ell_1\neq1$, the two arcs are no longer equivalent even before any trail is laid. The homing cost $\tilde{\kappa} \xt^2$ grows along the longer arc, so its returner resistance $R_i^-$ is larger. Returners split as $\Cc^-_i\propto 1/R^-_i$, hence the short arc receives more returners, more marking and hence more foragers. Length therefore acts as a fixed bias in favor of the short arc and trail feedback, in contrast, amplifies whichever arc is already better marked. The bias turns the pitchfork of the symmetric bridge into a single S-shaped branch and away from $r=1$ the short-trail strengthens (shown in Fig.~\ref{fig:bridge}(e), dynamics in SM Video 3). When $\asym>0$, it implies that the shorter arc~1 for $r>1$ carries more traffic and the long-arc state, by contrast, survives up to a fold. Which of the bistable states the colony occupies depends on its history. So a colony already committed to the long arc remains there and physically, the long arc survives only while the amplification of an established trail outweighs the geometric bias towards the short one; once the length difference is too large, the long-arc state is lost and the colony switches, as observed in~\cite{Goss1989,Beckers1992}. A second pair of folds appears when arc~2 is the shorter one, $0.919 < r < 0.954$ and all three states favor the short arc but with differences only on how strongly they select it.\\

\noindent\textit{Conclusion}---Our model of collective decision making in ants integrates the mathematical framework of optimal control with animal behavior in a manner that provides intuition for how macroscopic observations can be interpreted through the lens of value function/reward. By coupling the dynamics of outbound foragers to successful returners mediated by value field/pheromone, we show how collectives may choose between different trails. Further, the emergence of an Ohm's law for the steady-state ant density currents, along with the negative differential resistance driven instability, brings foraging task in line with non-linear resistor networks. While the reward structure and the behavioral rules are task specific, how collectives modulate their dynamics across tasks is a natural next question, which we will explore in our future work.

\begin{acknowledgments}
We thank Prof. L Mahadevan and Dr. Vishaal Krishnan for various discussions related to insect navigation and optimal control. We also thank the INTERFACE lab members for their inputs and suggestions. We acknowledge partial financial support from IIT Madras and ANRF-ECRG/2024/003341/ENS.\\
\end{acknowledgments}

\hskip -4mm {\bf Data availability}
\noindent The numerical solvers used in this study are available at \href{https://doi.org/10.5281/zenodo.23831931}{\texttt{Zenodo}}.

\hskip -4mm {\bf AI Assistance Statement} 
Assistance by Claude Opus was partially used for the following aspects of the work: developing the auto-07p solver, numerical solution of the non-linear HJB equations, verifying the theoretical findings, detecting typos, improving figure layout/design, improving grammar and language style within the manuscript.

\bibliographystyle{apsrev4-1}
\bibliography{references}

\clearpage
\onecolumngrid
\setcounter{section}{0}
\setcounter{equation}{0}
\setcounter{figure}{0}
\setcounter{table}{0}
\renewcommand{\thesection}{S\Roman{section}}
\renewcommand{\theequation}{S\arabic{equation}}
\renewcommand{\thefigure}{S\arabic{figure}}
\renewcommand{\thetable}{S\arabic{table}}

\begin{center}
{\large\bfseries Supplemental Material}\\[4pt]
\emph{Emergent stigmergy and pheromone-guided decisions in a foraging collective}
\end{center}
\vspace{6pt}

\renewcommand{\theHsection}{S\Roman{section}}
\renewcommand{\theHequation}{S\arabic{equation}}
\renewcommand{\theHfigure}{S\arabic{figure}}
\renewcommand{\theHtable}{S\arabic{table}}

\section{Notation}%
\label{sm:notation}

Table~\ref{tab:notation} below lists the symbols used in the article. Dimensional quantities carry no decoration, their scaled counterparts carry a tilde, and the scaling that relates them is given in Sec.~\ref{sm:onedim}.

\begin{table}[H]
\centering
\renewcommand{\arraystretch}{1.18}
\begin{tabular}{llll}
\hline\hline
Symbol & Definition & Non-dimensional form & Scaling\\
\hline
\multicolumn{4}{l}{\emph{State and geometry}}\\
$x$ & position, or arclength from the nest along an arc & $\xt$ & $\xt=x/L$\\
$\Om,\ON,\OF$ & domain, nest region, food region & --- & ---\\
$L$, $\ell_i$ & trail length, length of arc $i$ & $L_i$ & $L_1=1$, $L_2=r=\ell_2/\ell_1$\\
$S_t\in\{+,-\}$ & behavioral state, outbound or returning & --- & ---\\
\hline
\multicolumn{4}{l}{\emph{Parameters of the control problem}}\\
$D$ & diffusivity & $\varepsilon$ & $\varepsilon=D/\lambda L^{2}$\\
$\beta$ & mobility; running control cost is $|\bm u|^{2}/2\beta$ & --- & ---\\
$\lambda$ & Bellman discount rate, also the evaporation rate & --- & ---\\
$\alpha$, $\alpha_-$ & reward weight for marks, outbound and returning & $\sigma_-$ & $\sigma_-=\alpha_-/\alpha$\\
$r_F,r_N$ & conversion rewards at food and nest & $g_F,g_N$ & $g_{F,N}=\beta r_{F,N}/2D$\\
$U$ & homing potential paid by returners & $\Ut$ & $\Ut=\beta U/2D\lambda$\\
$\kappa$ & stiffness of $U=\tfrac12\kappa x^{2}$ & $\tilde\kappa$ & $\tilde\kappa=\beta\kappa L^{2}/4D\lambda$\\
$\Gamma_\pm$ & switching rates & --- & ---\\
$\sigma$ & deposition rate of outbound ants relative to returners & --- & ---\\
$\eta$ & cross-coupling & --- & $\eta=\sigma\sigma_-$\\
\hline
\multicolumn{4}{l}{\emph{Fields}}\\
$V^{\pm}$ & value functions & --- & ---\\
$Z,\Psi$ & Cole--Hopf variables & $w,v$ & $w=\ln Z$, $v=\ln\Psi$\\
 & $Z=e^{-\beta V^{+}/2D}$, $\Psi=e^{-\beta V^{-}/2D}$ & & \\
$\varrho_\pm$ & densities of outbound and returning ants & $\varrho_\pm$ & $\varrho_\pm=\alpha\beta\varrho_\pm/2D\lambda$\\
$m$ & mark density & $M$ & $M=\varrho_-+\sigma\varrho_+$\\
$\varphi_\pm$ & circuit potential & --- & $\varphi_+= \varepsilon\varrho_+/Z^{2}$, $\varphi_-=\varepsilon\varrho_-/\Psi^{2}$\\
$\phi$ & phenomenological pheromone concentration & --- & $\phi=Z-1$\\
\hline
\multicolumn{4}{l}{\emph{Integrated quantities}}\\
$\bm J_\pm$ & species currents & --- & $\bm J_\pm=2D\varrho_\pm\nabla\ln\chi_\pm-D\nabla\varrho_\pm$\\
$c$ & circulation on a single trail & $\Cc$ & $\Cc=\alpha\beta\,c/2D\lambda^{2}L$\\
$\Cc_i^{\pm}$ & arc currents on a network & --- & $\ctot=\sum_i \Cc_i^{+}=\sum_i \Cc_i^{-}$\\
$R^+_i,R^-_i$ & arc resistances & --- & $R^+_i=\int e^{-2w_i}\dd\xt$, $R^-_i=\int e^{-2v_i}\dd\xt$\\
$\Qtot$ & colony size & $\Qt$ & $\Qt=\alpha\beta\,\Qtot/2D\lambda L$\\
$\rmax$ & $\max_x\varrho_-$ & $\Theta$ & $\Theta=\alpha\beta\rmax/2D\lambda=\max_{\xt}\varrho_-$\\
$\asym$ & traffic asymmetry & --- & $\asym=(\Cc_1^{+}-\Cc_2^{+})/(\Cc_1^{+}+\Cc_2^{+})$\\
\hline\hline
\end{tabular}
\caption{Notation used throughout the article.}%
\label{tab:notation}
\end{table}

\section{Supplemental Videos}%
\noindent \href{https://drive.google.com/file/d/1UdqmCQCwL6ba6sPU85klbZPw-GBWQvXu/view?usp=sharing}{Video 1.} \textbf{Dynamics on a single trail} at $\varepsilon=0.05$, $g_F=6$, $g_N=2$, $\tilde\kappa=4$, in the camping ($\Qt=45$) and then the foraging ($\Qt=5$) phase, starting from the ant-free fields with all ants outbound in a Gaussian cloud at the nest. Space-time contours show $\varrho_\pm(x,t)$, $w(x,t)$ and $v(x,t)$, alongside the phase diagram of Fig.~\ref{fig:single}(b) with the operating point marked. The peak returner density $\Theta(t)=\max_x\varrho_-$ and the circulation $\ctot(t)$ relax through damped oscillations to their steady values (dotted).

\noindent \href{https://drive.google.com/file/d/1keKXVdPm-j3VsXK2xREdQ03IIsDwXVwA/view?usp=sharing}{Video 2.} \textbf{Dynamics on the symmetric double bridge} ($r=1$) at $\Qt=2.154$, below ($\sigma=\sigma_c-0.1$) and above ($\sigma=\sigma_c+0.1$) the pitchfork at $\sigma_c=0.8954$, with the other parameters as in Fig.~\ref{fig:bridge}. Space-time maps show $\varrho^{1,2}_\pm(x,t)$ on the two arcs, and the marker on the bifurcation diagram follows the instantaneous traffic asymmetry $\asym(t)$. Below $\sigma_c$ the initial bias decays to the symmetric state $\asym=0$; above it the colony commits to arc~1, with $\asym$ growing to the broken branch at $\asym\simeq0.60$ obtained from continuation (dotted).

\noindent \href{https://drive.google.com/file/d/1cUYPBS-u3CiSEUcLcBgMVMq0cBr9LlJz/view?usp=sharing}{Video 3.} \textbf{Dynamics on a double bridge with unequal arcs}, $r=\ell_2/\ell_1>1$, at $(\sigma-\sigma_c)=0.1$ and $\Qt=2.154$, with $\varrho^{1,2}_\pm(x,t)$, $\asym(t)$ and $\ctot(t)$ shown alongside the unfolding of Fig.~\ref{fig:bridge}(e). At $r=1.05$ only one steady state exists and the colony selects the shorter arc~1, with $\asym\simeq0.76$. At $r=1.01$, inside the bistable window, the outcome depends on the initial condition: a colony started on the longer arc~2 stays there ($\asym\simeq-0.49$), whereas an unbiased cloud at the nest selects the shorter arc ($\asym\simeq0.65$).
\section{Derivation of the coupled HJB system}%
\label{sm:hjb}

Consider an outbound ant at $X_t=x$, $S_t=+$ and over the interval $[t,t+\Delta t]$ the behavioral state is unchanged with probability $1-\Gamma_+(x)\Delta t$ and switches with probability $\Gamma_+(x)\Delta t$. Bellman's principle gives
\begin{equation}
\begin{aligned}
V^+(x,t)=\min_{\bm u_+}\Bigg\{
&\mathbb E_{x,+}^{\bm u_+}\!\left[\int_t^{t+\Delta t}e^{-\lambda(s-t)}
\left(\frac{|\bm u_+|^2}{2\beta}-\alpha\varrho_-(X_s,s)\right)\dd s\right]\\
&+e^{-\lambda\Delta t}\Big[(1-\Gamma_+(x)\Delta t)\,
\mathbb E_{x,+}^{\bm u_+}\!\left[V^+(X_{t+\Delta t},t+\Delta t)\right]\\
&\hspace{2.3cm}+\Gamma_+(x)\Delta t\,
\mathbb E_{x,+}^{\bm u_+}\!\left[V^-(X_{t+\Delta t},t+\Delta t)-r_F\right]\Big]\Bigg\}.
\end{aligned}
\label{eq:dpp}
\end{equation}
We can expand each ingredient to first order in $\Delta t$ and get:
\paragraph{Running cost.} The integrand is continuous and $e^{-\lambda(s-t)}=1+O(\Delta t)$ on the interval, so
\begin{equation}
\mathbb E_{x,+}^{\bm u_+}\!\left[\int_t^{t+\Delta t}e^{-\lambda(s-t)}
\left(\frac{|\bm u_+|^2}{2\beta}-\alpha\varrho_-\right)\dd s\right]
=\left(\frac{|\bm u_+(x,t)|^2}{2\beta}-\alpha\varrho_-(x,t)\right)\Delta t
+O(\Delta t).
\end{equation}
\paragraph{Continuation term.} It\^o's formula applied to $V^+(X_t,t)$ along Eq.~\eqref{eq:sde} gives
\begin{equation}
\dd V^+=\big(\partial_t V^++\bm u_+\!\cdot\!\nabla V^++D\nabla^2V^+\big)\dd t +\sqrt{2D}\,\nabla V^+\!\cdot\!\dd \bm W_t ,
\end{equation}
and the stochastic integral has zero mean, so
\begin{equation}
\mathbb E_{x,+}^{\bm u_+}\!\left[V^+(X_{t+\Delta t},t+\Delta t)\right]
=V^+(x,t)+\big(\partial_t V^++\bm u_+\!\cdot\!\nabla V^++D\nabla^2V^+\big)\Delta t
+O(\Delta t).
\end{equation}

\paragraph{Switching term.} This expectation is already multiplied by $\Gamma_+\Delta t$, so only its leading term $V^-(x,t)-r_F$ is retained. Substituting these three expansions into Eq.~\eqref{eq:dpp}, using $e^{-\lambda\Delta t}=1-\lambda\Delta t+O(\Delta t^2)$, canceling the $V^+(x,t)$ that appears on both sides, keeping terms of order $\Delta t$ and dividing by $\Delta t$ leaves
\begin{equation}
0=\min_{\bm u_+}\left\{\frac{|\bm u_+|^2}{2\beta}-\alpha\varrho_-
+\partial_t V^++\bm u_+\!\cdot\!\nabla V^++D\nabla^2V^+-\lambda V^+
+\Gamma_+(x)\big(V^--V^+-r_F\big)\right\}.
\label{eq:preinf}
\end{equation}
Only two terms in brackets depend on $\bm u_+$, and together they are
\begin{equation}
\frac{|\bm u_+|^2}{2\beta}+\bm u_+\!\cdot\!\nabla V^+
=\frac{1}{2\beta}\big|\bm u_++\beta\nabla V^+\big|^2
-\frac{\beta}{2}\big|\nabla V^+\big|^2 ,
\label{eq:completesquare}
\end{equation}
a perfect square plus a term independent of the control. The minimum is therefore attained at
\begin{equation}
\bm u^*_+=-\beta\nabla V^+ ,
\label{eq:ustar}
\end{equation}
where the first term on the right of Eq.~\eqref{eq:completesquare} vanishes and the bracket takes the value $-\dfrac{\beta}{2}|\nabla V^+|^2$. Hence
\begin{equation}
\partial_t V^++D\nabla^2V^+-\frac{\beta}{2}|\nabla V^+|^2-\lambda V^+
-\alpha\varrho_-+\Gamma_+(x)\big(V^--V^+-r_F\big)=0 .
\label{eq:hjbp}
\end{equation}

The returning state is treated identically and the running cost is now $|\bm u_-|^2/2\beta+U(x)$, the terminal payoff on switching is $V^+-r_N$, and the switching rate is $\Gamma_-$. Completing the square in the same way gives $\bm u^*_-=-\beta\nabla V^-$ and
\begin{equation}
\partial_t V^-+D\nabla^2V^--\frac{\beta}{2}|\nabla V^-|^2-\lambda V^-
+U(x)+\Gamma_-(x)\big(V^+-V^--r_N\big)=0 .
\label{eq:hjbm}
\end{equation}
Eqs.~\eqref{eq:hjbp} and~\eqref{eq:hjbm} are Eq.~\eqref{eq:hjb} of the article, with $f_+=-\alpha\varrho_-$ and $f_-=U$.

\section{Cole--Hopf transformation}%
\label{sm:colehopf}
We can write $V^+=-\dfrac{2D}{\beta}\ln Z$ with $Z>0$ and then
\begin{equation}
\partial_t V^+=-\frac{2D}{\beta}\frac{\partial_t Z}{Z},\qquad
\nabla V^+=-\frac{2D}{\beta}\frac{\nabla Z}{Z},\qquad
\nabla^2V^+=-\frac{2D}{\beta}\left(\frac{\nabla^2Z}{Z}-\frac{|\nabla Z|^2}{Z^2}\right),
\end{equation}
the last following from $\nabla\!\cdot\!(\nabla Z/Z)=\nabla^2Z/Z-|\nabla Z|^2/Z^2$. The two second-order terms of Eq.~\eqref{eq:hjbp} therefore combine as
\begin{equation}
D\nabla^2V^+-\frac{\beta}{2}|\nabla V^+|^2
=-\frac{2D^2}{\beta}\frac{\nabla^2Z}{Z}
+\frac{2D^2}{\beta}\frac{|\nabla Z|^2}{Z^2}
-\frac{\beta}{2}\cdot\frac{4D^2}{\beta^2}\frac{|\nabla Z|^2}{Z^2}
=-\frac{2D^2}{\beta}\frac{\nabla^2Z}{Z} .
\label{eq:quadcancel}
\end{equation}
The quadratic gradient term cancels identically and is why we assume the running cost as quadratic in $\bm u$ with the same coefficient structure as the noise. The optimal feedback in Eq.~\eqref{eq:ustar} becomes
\begin{equation}
\bm u^*_+=-\beta\nabla V^+=2D\frac{\nabla Z}{Z}=2D\nabla\ln Z ,
\end{equation}
which is Eq.~\eqref{eq:logsensing} of the article. Using $-\lambda V^+=\dfrac{2D\lambda}{\beta}\ln Z$ and $\Gamma_+(V^--V^+-r_F) =\Gamma_+\big[\dfrac{2D}{\beta}(\ln Z-\ln\Psi)-r_F\big]$, Eq.~\eqref{eq:hjbp} becomes
\begin{equation}
-\frac{2D}{\beta}\frac{\partial_t Z}{Z}-\frac{2D^2}{\beta}\frac{\nabla^2Z}{Z}
+\frac{2D\lambda}{\beta}\ln Z-\alpha\varrho_-
+\Gamma_+\Big[\frac{2D}{\beta}(\ln Z-\ln\Psi)-r_F\Big]=0 .
\end{equation}
Multiplying throughout by $-\beta Z/2D$ and rearranging,
\begin{equation}
\partial_t Z=-\Big(D\nabla^2Z+\frac{\alpha\beta}{2D}\varrho_-Z-\lambda Z\ln Z\Big)-\Gamma_+Z\Big(-\ln Z+\ln\Psi+\frac{\beta r_F}{2D}\Big),
\end{equation}
and the identical computation with $V^-=-\dfrac{2D}{\beta}\ln\Psi$ gives
\begin{equation}
\partial_t\Psi=-\Big(D\nabla^2\Psi-\frac{\beta}{2D}U\Psi-\lambda \Psi\ln \Psi\Big)-\Gamma_-\Psi\Big(-\ln \Psi+\ln Z+\frac{\beta r_N}{2D}\Big),
\end{equation}
which are Eqs.~\eqref{eq:Zfull} and~\eqref{eq:Psifull}.

\paragraph{Logarithmic variables.} For stable numerical solutions, the equations are best solved in logarithmic variables with Cole-Hopf step undone. Substituting $w=\ln Z$ we get $\nabla Z=Z\nabla w$ and
\begin{equation}
\nabla^2Z=\nabla\!\cdot\!(Z\nabla w)=Z\big(\nabla^2w+|\nabla w|^2\big),
\qquad Z\ln Z=Zw,
\end{equation}
so dividing the steady bulk equation $D\nabla^2Z-\lambda Z\ln Z+\dfrac{\alpha\beta}{2D}\varrho_-Z=0$ by $Z>0$ gives
\begin{equation}
D\big(\nabla^2w+|\nabla w|^2\big)-\lambda w+\frac{\alpha\beta}{2D}\varrho_-=0,
\label{eq:whjb}
\end{equation}
which is the HJB equation for $V^+=-2Dw/\beta$ written out. The logarithmic terms disappear, the optimal drift is $2D\nabla w$, so boundedness of the policy is boundedness of $\nabla w$, and the field stays of order unity in the regime $Z\gg1$ where a Newton solve on $Z$ can diverge. All quadratures of the form $\int Z^{-2}$ are evaluated as $\int e^{-2w}$ with the maximum of $-2w$ factored out,
\begin{equation}
\int e^{-2w}\,\dd\xt=e^{-2w_{\min}}\int e^{-2(w-w_{\min})}\,\dd\xt,
\qquad w_{\min}=\min_{\xt}w,
\end{equation}
so that the integrand never exceeds unity.

\section{Fokker--Planck equations and boundary conditions}%
\label{sm:fpe}

To obtain the generator for the density fields, condition on the switching event over $[t,t+\Delta t]$ with $a\in\{+,-\}$ labeling the behavioral state, we can write for any function $f(Y_t)$:
\begin{equation}
\mathbb E[f(Y_{t+\Delta t})|Y_t=(x,a)]
=(1-\Gamma_a(x)\Delta t)\,\mathbb E[f(X_{t+\Delta t},a)]
+\Gamma_a(x)\Delta t\,\mathbb E[f(X_{t+\Delta t},-a)]+O(\Delta t).
\end{equation}
It\^o's formula gives $\mathbb E[f(X_{t+\Delta t},a)]=f(x,a)+\Delta t\,(\bm u_a\!\cdot\!\nabla f+D\nabla^2f)+O(\Delta t)$, while the second term already carries a factor $\Delta t$ and needs only $f(x,-a)+O(1)$. Subtracting $f(x,a)$ and dividing by $\Delta t$,
\begin{equation}
(\mathcal Lf)(x,a)=\bm u_a\!\cdot\!\nabla f+D\nabla^2f-\Gamma_a(x)f(x,a)+\Gamma_a(x)f(x,-a).
\label{eq:gen}
\end{equation}
Integrating the drift and diffusion terms by parts and discarding boundary terms, we get for $f= \varrho_a$
\begin{equation}
\partial_t\varrho_a=-\nabla\!\cdot\!(\bm u_a\varrho_a)+D\nabla^2\varrho_a
-\Gamma_a\varrho_a+\Gamma_{-a}\varrho_{-a},\qquad a\in\{+,-\}.
\end{equation}
Substituting the optimal feedback $\bm u_+=2D\nabla\ln Z$ and $\bm u_-=2D\nabla\ln\Psi$,
\begin{align}
\partial_t\varrho_+&=-2D\nabla\!\cdot\!\big(\varrho_+\nabla\ln Z\big)+D\nabla^2\varrho_+
-\Gamma_+\varrho_++\Gamma_-\varrho_-,\\
\partial_t\varrho_-&=-2D\nabla\!\cdot\!\big(\varrho_-\nabla\ln\Psi\big)+D\nabla^2\varrho_-
+\Gamma_+\varrho_+-\Gamma_-\varrho_- ,
\end{align}
which is Eq.~\eqref{eq:fpe}, with currents
\begin{equation}
\bm J_\pm=2D\varrho_\pm\nabla\ln\chi_\pm-D\nabla\varrho_\pm,
\qquad \chi_+=Z,\ \chi_-=\Psi .
\label{eq:currents}
\end{equation}

\paragraph{Zero-current solution.} Setting $\bm J_+=0$ in Eq.~\eqref{eq:currents} gives $\nabla\varrho_+/\varrho_+=2\nabla\ln Z$, that is $\nabla\ln\varrho_+=\nabla\ln Z^2$, so $\varrho_+=\text{const}\times Z^2$. Equivalently $\varrho_+\propto e^{-\beta V^+/D}$, a Gibbs distribution over the value function at temperature $D/\beta$. This is Eq.~\eqref{eq:boltzmann} of the article. It is the natural detailed-balance reference state and should be used to validate any numerical scheme in a regime with no circulation.

\paragraph{Boundary conditions.} On a bounded reflecting domain the wall passes no ants, but it does not separately annihilate each species. We then get
\begin{equation}
\partial_n Z=\partial_n\Psi=0,
\qquad
\big(\bm J_++\bm J_-\big)\!\cdot\!\bm n=0
\qquad\text{on }\partial\Om .
\label{eq:wallbc}
\end{equation}
The conditions $\bm J_\pm\!\cdot\!\bm n=0$, and therefore $\partial_n\varrho_\pm=0$, hold only on the part of $\partial\Om$ disjoint from $\partial\OF$ and $\partial\ON$. Where a conversion region is co-located with the wall, conversion turns $+$ into $-$ at the wall itself, so both species carry a nonzero normal current there while their sum still vanishes. Imposing the species-wise form on all of $\partial\Om$ would forbid exactly the geometry recommended in Sec.~\ref{sm:interface}. On an unbounded domain the conditions are $Z\to1$, $\Psi\to0$ and $\varrho_\pm\to0$ at infinity.

\section{Instantaneous-switching limit and interface conditions}%
\label{sm:interface}

Let $\Gamma_+\to\infty$ on $\OF$ and $\Gamma_-\to\infty$ on $\ON$. In the bulk the switching terms drop out of the value equations,
\begin{align}
D\nabla^2Z-\lambda Z\ln Z+\frac{\alpha\beta}{2D}\varrho_-Z&=0,
&&x\in\Om\setminus\OF,\\
D\nabla^2\Psi-\lambda \Psi\ln \Psi-\frac{\beta}{2D}U\Psi&=0,
&&x\in\Om\setminus\ON,
\end{align}
while inside the conversion regions the switching terms dominate and force their brackets to vanish, which clamps the values,
\begin{equation}
Z=\Psi\,e^{\beta r_F/2D}\ \ (x\in\OF),\qquad
\Psi=Z\,e^{\beta r_N/2D}\ \ (x\in\ON),
\end{equation}
with $[Z]=[\Psi]=0$ and $[\partial_n Z],[\partial_n\Psi]\neq0$ across $\partial\OF$ and $\partial\ON$, where $[\cdot]$ denotes the jump.

\subsection*{Flux conditions}

\paragraph{Total current.} Switching conserves the total number of ants, so adding the two Fokker--Planck equations gives $\partial_t(\varrho_++\varrho_-)+\nabla\!\cdot\!(\bm J_++\bm J_-)=0$ exactly, for any $\Gamma_\pm$. In steady state $\nabla\!\cdot\!\bm J_{\rm tot}=0$ throughout $\Om$, hence
\begin{equation}
\big[\bm J_+\!\cdot\!\bm n\big]+\big[\bm J_-\!\cdot\!\bm n\big]=0
\qquad\text{on }\partial\OF\text{ and }\partial\ON.
\label{eq:totflux}
\end{equation}

\paragraph{Individual currents.} To estimate the individual currents, we integrate $\partial_t\varrho_+=-\nabla\!\cdot\!\bm J_+-\Gamma_+\varrho_++\Gamma_-\varrho_-$ across a layer of thickness $2\ell$ around $\partial\OF$ and let $\ell\to0$. The term $\Gamma_-\varrho_-$ stays bounded and contributes nothing, while $\Gamma_+\varrho_+$ concentrates, so $[\bm J_+\!\cdot\!\bm n]=-j_F$ with
\begin{equation}
j_F(x)=\lim_{\ell\to0}\int_{-\ell}^{\ell}\Gamma_+\varrho_+\,\dd x_\perp\ \ge\ 0
\end{equation}
the surface conversion current density. The same integration for $\varrho_-$ gives $[\bm J_-\!\cdot\!\bm n]=+j_F$, consistent with Eq.~\eqref{eq:totflux}. The correct distributional form of the steady transport equations is therefore
\begin{equation}
\nabla\!\cdot\!\bm J_+=-j_F\,\delta_{\partial\OF}+j_N\,\delta_{\partial\ON},
\qquad
\nabla\!\cdot\!\bm J_-=+j_F\,\delta_{\partial\OF}-j_N\,\delta_{\partial\ON},
\label{eq:distributional}
\end{equation}
where $\delta_{\partial\OF}$ is the delta-function on $\partial\OF$ and $j_F,j_N\ge0$ are determined as part of the solution. Because $\varrho_+\equiv0$ inside $\OF$, and hence $\bm J_+=0$ there, the conversion rate is fixed by the arriving flux, $j_F=\bm J_+\!\cdot\!\bm n|_{\rm out}$, and similarly $j_N=\bm J_-\!\cdot\!\bm n|_{\rm out}$. The circulation is
\begin{equation}
c=\int_{\partial\OF}j_F\,\dd S=\int_{\partial\ON}j_N\,\dd S.
\end{equation}

Eq.~\eqref{eq:distributional} shows that $\partial\ON$ and $\partial\OF$ are the only sources and sinks of each species, since away from them $\nabla\!\cdot\!\bm J_\pm=0$. It is worth recording exactly when this forbids a circulating state.

\paragraph{Criterion.} A bare reflecting wall and a Dirichlet zero for the same species cannot bound a connected region containing no source for that species. Suppose $\varrho_+$ has no source on such a region, then $\bm J_+$ is divergence-free with vanishing normal component on the bare wall, so $\bm J_+\equiv0$, and the zero-current solution $\varrho_+\propto Z^2$ applies. Since $Z\ge1$ by Sec.~\ref{sm:positivity}, the factor $Z^2$ cannot vanish, so the Dirichlet condition $\varrho_+=0$ at the other end sets the prefactor to zero and $\varrho_+\equiv0$ identically.

\paragraph{Illustration in one dimension.} Take $\Om=[0,L]$ with a bare wall at $x=0$ and the food at $x=L$, and delete the nest source. The criterion applies to $[0,L)$ immediately, and there is no circulating steady state at all. This is not a statement about $x=0$ being distinguished, nor an asymmetry between the two species: the corresponding statement for $\varrho_-$ is never invoked, because a conversion region co-located with a wall satisfies only Eq.~\eqref{eq:wallbc} and does not force $\bm J_-$ to vanish there.

\paragraph{Circuit analogy.} The nest is an electromotive force that pumps $+$ and sinks $-$, and the food does the reverse. Current flows only on paths connecting them and any part of the domain not on such a path is a dead carrying zero current, with each species in its zero-flux Gibbs state. Deleting the nest removes the only source of electromotive force in the circuit.

\paragraph{Resolution.} The nest need not be in the interior of the domain. Placing the nest and the food at the two ends of the segment, each co-located with a reflecting wall, gives
\begin{equation}
Z'(0)=0,\quad \Psi(0)=Z(0)e^{\beta r_N/2D},\qquad
\Psi'(L)=0,\quad Z(L)=\Psi(L)e^{\beta r_F/2D},
\label{eq:sym1d}
\end{equation}
two conditions on each field, clamped at opposite ends, with no dead stubs anywhere. The structure is mirror-symmetric: $Z$ carries a Neumann condition at the nest end and a Dirichlet at the food end, and $\Psi$ the reverse. The two Neumann conditions are the walls and the two Dirichlet conditions are the conversion rewards.

\subsection*{\texorpdfstring{Dirichlet conditions $\varrho_\pm=0$}{Status of the Dirichlet conditions rho=0}}

The conditions $\varrho_+|_{\partial\OF}=0$ and $\varrho_-|_{\partial\ON}=0$ are the $\Gamma\to\infty$ limit of an interior layer of width $\ell_\Gamma=\sqrt{D/\Gamma}$, not exact boundary conditions at finite $\Gamma$. Within the layer $\varrho_+$ decays as $e^{-x_\perp/\ell_\Gamma}$, and at finite $\Gamma$ the correct condition is the Robin form $\bm J_+\!\cdot\!\bm n=\sqrt{D\Gamma_+}\,\varrho_+$. However, we use the sharp-interface form throughout and the $O(\ell_\Gamma)$ correction to $\Cc$ has not been computed, and the Neumann conditions $Z'(0)=\Psi'(L)=0$ are likewise the zero-width limit of nest and food regions of finite extent.

\section{Positivity of the Cole--Hopf variable}%
\label{sm:positivity}

\paragraph{Finiteness of $V^+$.} $Z=e^{-\beta V^+/2D}$ is strictly positive wherever $V^+$ is finite, which is required before $\ln Z$ can be written at all and is assumed throughout. On a bounded domain $Z$ is continuous and on an unbounded domain the condition $Z\to1$ at infinity means that if $\min Z<1$ then the minimum is attained within the domain.

If the viability condition in Eq.~\eqref{eq:theta} holds, then we can show that $Z(x)\ge1$ for all $x\in\Om\setminus\OF$. Suppose $\min_{\Om\setminus\OF}Z<1$, attained at $x_*$ and $x_*$ is interior then $\nabla^2Z(x_*)\ge0$. The governing equation gives
\begin{equation}
D\nabla^2Z=\lambda Z\ln Z-\frac{\alpha\beta}{2D}\varrho_-Z .
\end{equation}
Since $0<Z(x_*)<1$ we have $\ln Z(x_*)<0$ and hence $Z\ln Z<0$ strictly, while $\varrho_-\ge0$ makes the second term non-positive. Therefore $\nabla^2Z(x_*)<0$, a contradiction.

\paragraph{Food interface.} By the viability condition Eq.~\eqref{eq:theta}, $Z(x_F)>e^{\Theta}>1$, so the minimum cannot lie there, hence $Z(x)\ge1$ on $\Om\setminus\OF$.

From here we see that since $\ln Z\ge0$, so $-\lambda Z\ln Z\le0$ everywhere and the discount term is a sink. Without this, regions with $Z<1$ would have $\ln Z<0$ and the same term would act as a spontaneous source, which breaks the pheromone analogy and also numerically unstable. Further, the resistivity satisfies $0<Z^{-2}\le1$, so every arc resistance $R^+_i$ of Sec.~\ref{sm:graph} is limited by the arc length: a trail can only lower the resistance below its bare diffusive value, never raise it.

\section{Viability conditions}%
\label{sm:conditions}

\paragraph{Condition 1: food preferred over nest.} Since $\bm u^*_+=-\beta\nabla V^+$, outbound ants descend $V^+$, and food is the globally preferred destination when
\begin{equation}
Z(x_F)>Z(x_N)\iff V^+(x_F)<V^+(x_N).
\end{equation}
Gradient descent is local, so this global comparison is necessary but not sufficient. A local minima of $V^+$ away form the food could trap the descent which is exactly the camping phase.

\paragraph{Condition 2: food worth reaching.} The transmission condition is $V^+(x_F)=V^-(x_F)-r_F$, so requiring $V^+(x_F)<0$ gives $r_F>V^-(x_F)$, that is $Z(x_F)>1$.

\paragraph{Condition 3: terminal reward dominates the shaping reward.} Let $\rmax=\max_x\varrho_-$. For an outbound ant that reaches food at the random time $\tau$, the cost of an arbitrary admissible policy is
\begin{equation}
\mathcal K[\bm u_+,\tau]=\mathbb E\left[\int_0^\tau e^{-\lambda t}
\left(\frac{|\bm u_+|^2}{2\beta}-\alpha\varrho_-(X_t)\right)\dd t\right]
+\mathbb E\left[e^{-\lambda\tau}\big(V^-(x_F)-r_F\big)\right].
\end{equation}
The control cost is non-negative and $\varrho_-\le\rmax$ pointwise, so the integrand is bounded below by $-\alpha\rmax e^{-\lambda t}$. Carrying out the resulting integral,
\begin{equation}
\int_0^\tau\!\!-\alpha\rmax e^{-\lambda t}\,\dd t
=-\frac{\alpha\rmax}{\lambda}\big(1-e^{-\lambda\tau}\big),
\end{equation}
and therefore, for deterministic $\tau$,
\begin{equation}
\mathcal K[\bm u_+,\tau]\ \ge\ \mathcal K_{\rm lb}(\tau)
:=-\frac{\alpha \rmax}{\lambda}\big(1-e^{-\lambda\tau}\big)
+e^{-\lambda\tau}\big(V^-(x_F)-r_F\big).
\end{equation}
The bound holds for every admissible control and every switching time, so the correct conclusion is $V^+(x)\ge\min_{\tau>0}\mathcal K_{\rm lb}(\tau)$ and not $V^+(x)\ge\mathcal K_{\rm lb}(\tau)$ for one particular $\tau$.

Compare with the idealised cost of camping at the pheromone maximum forever,
\begin{equation}
\mathcal K_{\rm camp}=-\alpha\rmax\int_0^\infty e^{-\lambda t}\,\dd t
=-\frac{\alpha \rmax}{\lambda},
\end{equation}
which omits the control cost of resisting diffusion and is therefore a conservative benchmark. Subtracting,
\begin{equation}
\mathcal K_{\rm lb}(\tau)-\mathcal K_{\rm camp}=e^{-\lambda\tau}
\left[\frac{\alpha \rmax}{\lambda}+V^-(x_F)-r_F\right],
\end{equation}
and since $e^{-\lambda\tau}>0$ the sign of the bracket is independent of $\tau$. Hence if $r_F-V^-(x_F)\le\alpha \rmax/\lambda$ then $\min_\tau \mathcal K_{\rm lb}\ge \mathcal K_{\rm camp}$ and no food-reaching strategy beats camping. The contrapositive is the necessary condition
\begin{equation}
r_F-V^-(x_F)>\frac{\alpha \rmax}{\lambda}
\iff -V^+(x_F)>\frac{\alpha \rmax}{\lambda}
\iff \ln Z(x_F)>\Theta\equiv\frac{\alpha\beta \rmax}{2D\lambda},
\end{equation}
the last step using $V^+=-\dfrac{2D}{\beta}\ln Z$. Since $\Theta>0$, Condition~3 implies Condition~2. Both sides of this inequality are outputs of the converged solution, so the condition is tested rather than imposed. Sec.~\ref{sm:onedim} records $\Theta$ and $w(1)=\ln Z(x_F)$ at every point of the sweep. Because the benchmark $\mathcal K_{\rm camp}$ neglects the control cost of staying put, the condition is necessary and not sufficient, and the measured onset of camping lies above the bound; Fig.~\ref{fig:phase} shows the gap.

\section{Weak-trail limit and its applicability}%
\label{sm:weaktrail}
For completeness we record the linearization that recovers the canonical pheromone equation, and expand on its validity. When we set $\phi=Z-1\ge0$ in Eq.~\eqref{eq:pheromone_exact} and expand for $\phi\ll1$ and $\varrho_-\ll1$, using
\begin{equation}
(1+\phi)\ln(1+\phi)=\phi+\tfrac12\phi^2+O(\phi^3),\qquad
\varrho_-(1+\phi)=\varrho_-+\varrho_-\phi,
\end{equation}
and discard the second-order terms $\tfrac12\phi^2$ and $\varrho_-\phi$, we get
\begin{equation}
\partial_t\phi=D\nabla^2\phi+\frac{\alpha\beta}{2D}\varrho_--\lambda\phi ,
\label{eq:canonical}
\end{equation}
which is the canonical pheromone reaction--diffusion equation, with the Bellman discount rate $\lambda$ appearing as the evaporation rate.

\paragraph{Validity.} Eq.~\eqref{eq:canonical} requires $\phi\ll1$, that is $\ln Z\ll1$. Condition~3 of Sec.~\ref{sm:conditions} requires $\ln Z(x_F)>\Theta$. The two are compatible only if $\Theta\ll1$, that is $\alpha\beta\rmax\ll2D\lambda$: weak shaping reward, cheap control, or fast evaporation. In this regime the trail is close to diffusive search and whenever the trail is strong enough the linearization fails. Thus this approximation is valid far from the food-source or when the reward is weak or when foraging fails. We therefore retain Eq.~\eqref{eq:pheromone_exact} in full and regard Eq.~\eqref{eq:canonical} as a control case rather than as the physical equation.

\paragraph{Interpretation of full equation.} We can write forward propagating Eq.~\eqref{eq:pheromone_exact} as $\partial_t Z=D\nabla^2Z+k_{\rm dep}-k_{\rm evap}Z$ with $k_{\rm dep}=\dfrac{\alpha\beta}{2D}\varrho_-Z$ and $k_{\rm evap}=\lambda\ln Z$. Deposition is autocatalytic in the existing trail, and the effective evaporation rate grows logarithmically with concentration. The circuit reduction sharpens the distinction here as well. In the linear theory the resistivity would be $(1+\phi)^{-2}\simeq1-2\phi$, so a trail lowers the resistance by a fraction of itself. In the exact theory it is $e^{-2w}$, which is exponentially small. Selection strength in a branched network is controlled by the ratio of arc resistances, so the difference between the two is the difference between a weak preference and an exponentially strong one.

\section{One-dimensional trail}%
\label{sm:onedim}
The domain is $[0,L]$, with a point nest at $x=0$ co-located with a reflecting wall, and the food at $x=L$ likewise co-located with a reflecting wall. Both fields are defined on the whole of $[0,L]$ and the current is nonzero everywhere, $J_+=+c$ and $J_-=-c$ on $(0,L)$, the sign convention being that returners travel towards the nest. There are no quiescent regions and the steady value equations are
\begin{align}
D\,Z'' - \lambda Z\ln Z + \frac{\alpha\beta}{2D}\varrho_- Z &= 0,
&& x\in[0,L), 
\label{eq:Zdim}\\
D\,\Psi'' - \lambda\Psi\ln\Psi - \frac{\beta}{2D}U(x)\,\Psi &= 0,
&& x\in(0,L], 
\label{eq:Psidim}
\end{align}
with currents $J_+ = 2D\varrho_+(\ln Z)' - D\varrho_+'$ and $J_- = 2D\varrho_-(\ln\Psi)' - D\varrho_-'$, the boundary conditions in Eq.~\eqref{eq:sym1d} on the value fields, and
\begin{equation}
\varrho_+(L)=0,\qquad \varrho_-(0)=0,\qquad
\int_0^L(\varrho_++\varrho_-)\,\dd x=\Qtot
\label{eq:densbc}
\end{equation}
on the densities. At both walls $J_++J_-=0$ is satisfied identically.

Since $Z=e^{-\beta V^+/2D}$ with $Z\ge1$, we have $\ln Z\ge0$. The returner's value $V^-$ is a cost, so $\Psi<1$ and $\ln\Psi<0$ in the bulk. At the nest the returner collects $r_N$, so $\Psi(0)>1$. At the food, $Z(L)>1$ requires $r_F$ to exceed the residual cost $V^-(L)$, which is Condition~2 of Sec.~\ref{sm:conditions}.

\subsection*{Non-dimensionalisation}

We scale length by the domain size, $\xt=x/L$, so that $\partial_x^2=L^{-2}\partial_{\xt}^2$. Eq.~\eqref{eq:Zdim} becomes
\begin{equation}
\frac{D}{L^{2}}\,\partial^2_{\xt}Z-\lambda Z\ln Z
+\frac{\alpha\beta}{2D}\varrho_-Z=0 ,
\end{equation}
and dividing throughout by $\lambda$,
\begin{equation}
\underbrace{\frac{D}{\lambda L^{2}}}_{\displaystyle\varepsilon}\,Z''
-Z\ln Z+\underbrace{\frac{\alpha\beta}{2D\lambda}\varrho_-}_{\displaystyle\varrho_-}Z=0 .
\end{equation}
The same scaling on Eq.~\eqref{eq:Psidim} gives $\Ut=\beta U/2D\lambda$. Hence
\begin{equation}
\varepsilon Z'' - Z\ln Z + \varrho_-Z = 0,\qquad
\varepsilon \Psi'' - \Psi\ln\Psi - \Ut\,\Psi = 0,
\label{eq:nondim}
\end{equation}
where $'$ denotes $\dd/\dd\xt$. For the harmonic homing potential $U=\dfrac12\kappa x^{2}=\dfrac12\kappa L^{2}\xt^{2}$ one has $\Ut=\tilde\kappa\,\xt^{2}$ with $\tilde\kappa=\beta\kappa L^{2}/4D\lambda$.

The consequence of this scaling is that
\begin{equation}
\Theta=\frac{\alpha\beta\rmax}{2D\lambda}=\max_{\xt}\varrho_-(\xt),
\end{equation}
so $\Theta$ is not an input parameter and it has to be found out from the converged solution. Applying the same scaling to the current and using $\tfrac{\alpha\beta}{2D\lambda}\varrho_\pm=\varrho_\pm$,
\begin{equation}
\frac{\alpha\beta}{2D\lambda}J_+
=\frac{\alpha\beta}{2D\lambda}\cdot\frac{D}{L}
\Big(2\varrho_+(\ln Z)'-\varrho_+'\Big)
=\frac{D}{L}\Big(2\varrho_+w'-\varrho_+'\Big).
\end{equation}
Setting $J_+=c$ and dividing by $\lambda L$,
\begin{equation}
\varepsilon\Big(2\varrho_\pm(\ln\chi_\pm)' - \varrho_\pm'\Big) = \pm \Cc,
\qquad \chi_+=Z,\ \ \chi_-=\Psi,
\qquad
\Cc \equiv \frac{\alpha\beta\,c}{2D\lambda^{2}L} .
\label{eq:cdef}
\end{equation}
The population constraint scales the same way and multiplying Eq.~\eqref{eq:densbc} by $\alpha\beta/2D\lambda$ and substituting $\dd x=L\dd\xt$,
\begin{equation}
\int_0^1\big(\varrho_++\varrho_-\big)\,\dd\xt = \Qt,
\qquad
\Qt \equiv \frac{\alpha\beta\,\Qtot}{2D\lambda L}.
\label{eq:popcon}
\end{equation}

\noindent \textit{Control parameters.} We have five non-dimensional numbers, plus the choice of homing potential: $\varepsilon, g_F, g_N, \Qt$, and form of $\Ut$. The natural sweep is in total population $\Qt$, equivalently in $\alpha$; $\Theta$ then can be estimated from the solution.

\subsection*{Logarithmic variables}

Substituting $w=\ln Z$ and $v=\ln\Psi$ into Eq.~\eqref{eq:nondim}, and using $Z''=Z(w''+w'^2)$ and $Z\ln Z=Zw$ as in Sec.~\ref{sm:colehopf}, division by $Z>0$ and $\Psi>0$ gives
\begin{equation}
\varepsilon\big(w'' + w'^2\big) - w + \varrho_- = 0,
\qquad
\varepsilon\big(v'' + v'^2\big) - v - \Ut = 0 .
\label{eq:logeqs}
\end{equation}
The optimal drifts are $u_+ = (2D/L)\,w'$ and $u_-=(2D/L)\,v'$, so boundedness of the policy is boundedness of $w'$ and $v'$. Because $Z>0$, the boundary conditions in Eq.~\eqref{eq:sym1d} map one-to-one onto
\begin{equation}
w'(0)=0,\qquad v(0)=w(0)+g_N,\qquad v'(1)=0,\qquad w(1)=v(1)+g_F ,
\label{eq:logbcs}
\end{equation}
and the viability condition Eq.~\eqref{eq:theta} reads $w(1)>\Theta$.

\subsection*{Closed-form densities}

The solution to the density of outbound and returners involves no boundary-value problem and no shooting for a fixed $c$. We can define the scaled potential
\begin{equation}
\varphi_+(\xt)\;\equiv\;\varepsilon\,\varrho_+(\xt)\,e^{-2w(\xt)}
\;=\;\varepsilon\,\frac{\varrho_+(\xt)}{Z^{2}(\xt)} ,
\label{eq:potentialscaled}
\end{equation}
which is the $\varphi_+$ of Eq.~\eqref{eq:potential} multiplied by $\alpha\beta/2D\lambda^{2}L^{2}$. Differentiating and using the product rule,
\begin{equation}
\varphi_+' = \varepsilon e^{-2w}\big(\varrho_+' - 2\varrho_+w'\big)
= -\,e^{-2w}\cdot\varepsilon\big(2\varrho_+w'-\varrho_+'\big)
= -\,\Cc\,e^{-2w},
\label{eq:ohmscaled}
\end{equation}
the last step by Eq.~\eqref{eq:cdef}. The drift terms have cancelled identically; this is the first integral referred to in the article. Integrating from $\xt$ to $1$ and imposing the ground $\varphi_+(1)=0$, which follows from $\varrho_+(1)=0$, gives $\varphi_+(\xt)=\Cc\int_{\xt}^{1}e^{-2w}$, that is
\begin{equation}
\varrho_+(\xt) = \frac{\Cc}{\varepsilon}\,e^{2w(\xt)}
\int_{\xt}^{1}e^{-2w(\tilde y)}\,\dd\tilde y .
\label{eq:pplus}
\end{equation}
A similar calculation for the returners, with $\varphi_-=\varepsilon\varrho_-e^{-2v}$, $\varphi_-'=+\Cc\,e^{-2v}$ and the ground $\varphi_-(0)=0$ from $\varrho_-(0)=0$, gives
\begin{equation}
\varrho_-(\xt) = \frac{\Cc}{\varepsilon}\,e^{2v(\xt)}
\int_{0}^{\xt}e^{-2v(\tilde y)}\,\dd\tilde y .
\label{eq:pminus}
\end{equation}
Substituting Eqs.~\eqref{eq:pplus} and~\eqref{eq:pminus} into the population constraint Eq.~\eqref{eq:popcon} and solving for $\Cc$ gives the circulation,
\begin{equation}
\Cc = \varepsilon\,\Qt \bigg/
\left[
\int_0^1 e^{2w(\xt)}\!\!\int_{\xt}^{1}\!\!e^{-2w}\,\dd\tilde y\,\dd\xt
\;+\;
\int_{0}^1 e^{2v(\xt)}\!\int_{0}^{\xt}e^{-2v}\,\dd\tilde y\,\dd\xt
\right].
\label{eq:cquad}
\end{equation}
Each integrand here is evaluated in logarithmic variables with the maximum of the exponent factored out, as described in Sec.~\ref{sm:colehopf}.

\subsection*{Camping phase}

In the camping phase $w$ has an interior maximum at $\xt_*$, where $w'(\xt_*)=0$ and $w''(\xt_*)<0$. Evaluating Eq.~\eqref{eq:logeqs} there gives $\varepsilon w''(\xt_*)=w(\xt_*)-\varrho_-(\xt_*)$, so a maximum requires $\varrho_-(\xt_*)>w(\xt_*)$: deposition exceeds evaporation at the peak.

\paragraph*{Quadratic approximation.} Since $\varrho_+\propto e^{2w}$, the outbound density is carried only by the neighbourhood of $\xt_*$ where $w$ is within order one of its maximum, of width $\sim|w''|^{-1/2}$. We therefore expand $w$ to second order there,
\begin{equation}
w(\xt)\simeq w(\xt_*)-\tfrac12|w''(\xt_*)|\,(\xt-\xt_*)^2 .
\label{eq:quadw}
\end{equation}
This is a local approximation of Eq.~\eqref{eq:logeqs}, accurate if the cubic term is small over the same width. Differentiating Eq.~\eqref{eq:logeqs} at $\xt_*$ gives $w'''(\xt_*)=-\varrho_-'(\xt_*)/\varepsilon$, so the cubic term is small when $\varrho_-$ changes little across the camp. This is the case in the numerical solutions, where $\varrho_-$ is smooth and nearly flat around the peak of $w$ (Fig.~\ref{fig:single}(c)).

\paragraph*{Gaussian density.} The integral $\int_{\xt}^{1}e^{-2w}\dd\tilde y$ in Eq.~\eqref{eq:pplus} is dominated by the food end, where $w$ is smallest, and so is nearly constant across the camp. Then $\varphi_+$ is constant there and $\varrho_+$ is locally the Gibbs state Eq.~\eqref{eq:boltzmann}, $\varrho_+\propto Z^2$. With Eq.~\eqref{eq:quadw} this is a Gaussian,
\begin{equation}
\varrho_+(\xt)\simeq\varrho_+(\xt_*)\,e^{-|w''(\xt_*)|(\xt-\xt_*)^2},
\label{eq:gausscamp}
\end{equation}
of variance $1/2|w''|=\varepsilon/2(\varrho_--w)$ at $\xt_*$. The circulation follows from Eq.~\eqref{eq:pplus} at $\xt_*$,
\begin{equation}
\Cc\simeq\varepsilon\,\varrho_+(\xt_*)\,e^{-2w(\xt_*)}\bigg/\int_{\xt_*}^{1}e^{-2w}\dd\tilde y ,
\label{eq:ccamp}
\end{equation}
which becomes exponentially small at $\xt_*$.

\paragraph*{Zero food reward.} For $g_F=0$ the clamp Eq.~\eqref{eq:logbcs} gives $w(1)=v(1)$, and for $\varepsilon\ll1$, $v\simeq-\tilde\kappa\,\xt^2$ away from the boundary layers, so $w(1)\simeq-\tilde\kappa<0<\Theta$. By Condition~3 of Sec.~\ref{sm:conditions}, no food-reaching policy can have a lower cost than camping, and the system is in the camping phase.

\subsection*{Phase diagram across parameters}

\begin{figure}[H]
\centering
\includegraphics[width=\textwidth]{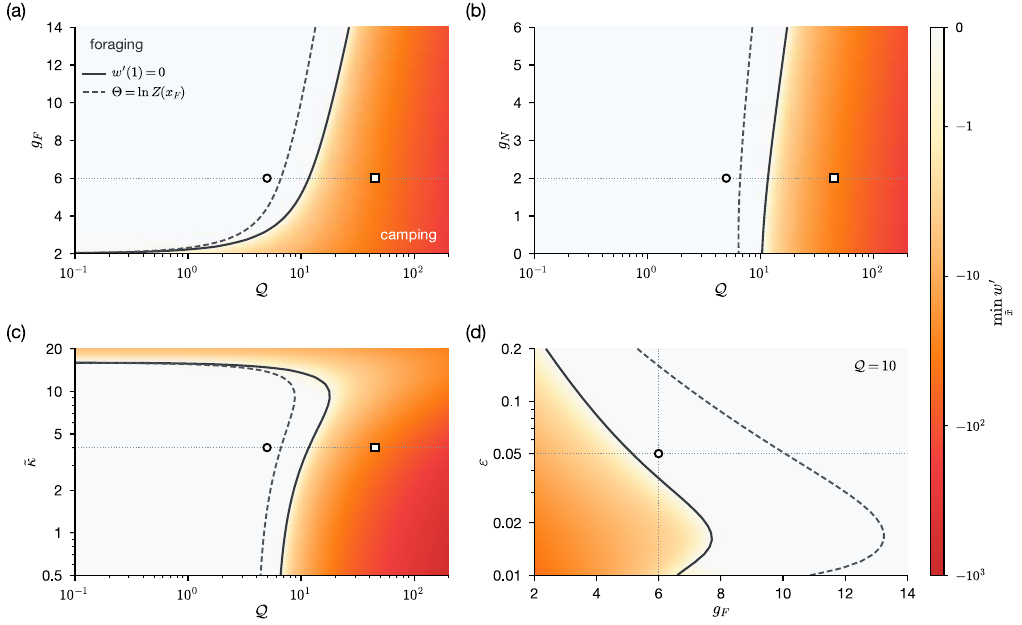}
\caption{Phase diagrams of single-trail foraging across parameters. Smallest outbound drift $\min_{\xt}w'$ in the planes (a) $(\Qt,g_F)$, (b) $(\Qt,g_N)$, (c) $(\Qt,\tilde\kappa)$ and (d) $(g_F,\varepsilon)$. Solid: camping onset $w'(1)=0$; dashed: $\Theta=\ln Z(x_F)$; markers are Circle: $\Qt=5$, and Square: $\Qt=45$. Parameters not on an axis are $\varepsilon=0.05$, $g_F=6$, $g_N=2$, $\tilde\kappa=4$ and $\Qt=10$.}
\label{fig:phase}
\end{figure}

Figure~\ref{fig:single}(b) shows one slice of the control space spanned by $\varepsilon$, $g_F$, $g_N$, $\Qt$ and the form of $\Ut$. Figure~\ref{fig:phase} extends it to the other directions, using the same order parameter $\min_{\xt}w'$ of the logarithmic variables Eq.~\eqref{eq:logeqs}: $\Qt$ against $g_F$, $g_N$ and $\tilde\kappa$, and $g_F$ against $\varepsilon$. The solid line is the measured camping onset $w'(1)=0$ and the dashed line is the bound Eq.~\eqref{eq:theta}, $\Theta=\ln Z(x_F)$. The two do not coincide on any map.

\subsection*{Time dynamics of the single trail}

The steady state of this section is the fixed point of the forward-in-time system. The fields obey
\begin{equation}
\partial_{\tilde{t}} w=\varepsilon\big(w''+w'^2\big)-w+\varrho_- ,
\qquad
\partial_{\tilde{t}} v=\varepsilon\big(v''+v'^2\big)-v-\Ut ,
\label{eq:dynfields}
\end{equation}
with the conditions in Eq.~\eqref{eq:logbcs} imposed at every time, and the densities obey
\begin{equation}
\partial_{\tilde{t}}\varrho_\pm=-\,\partial_{\xt}\Big[\varepsilon\Big(2\varrho_\pm(\ln\chi_\pm)'-\varrho_\pm'\Big)\Big],
\qquad \chi_+=Z,\ \ \chi_-=\Psi,
\label{eq:dynrho}
\end{equation}
with $\varrho_+(1)=\varrho_-(0)=0$. Outbound ants arriving at the food become returners there and returners arriving at the nest become outbound there, so $\int_0^1(\varrho_++\varrho_-)\,\dd\xt=\Qt$ is conserved at all times. At stationarity, Eqs.~\eqref{eq:dynfields}--\eqref{eq:dynrho} reduce to Eqs.~\eqref{eq:logeqs} and~\eqref{eq:cdef}.

We start from an ant-free steady fields, the solution of Eq.~\eqref{eq:logeqs} at $\Qt=0$ to solve for the intrinsic desirability fields of the food and nest, and then introduce outbound ants in a Gaussian cloud of width $0.02$ at the nest.

\begin{figure}[H]
\centering
\includegraphics[width=\textwidth]{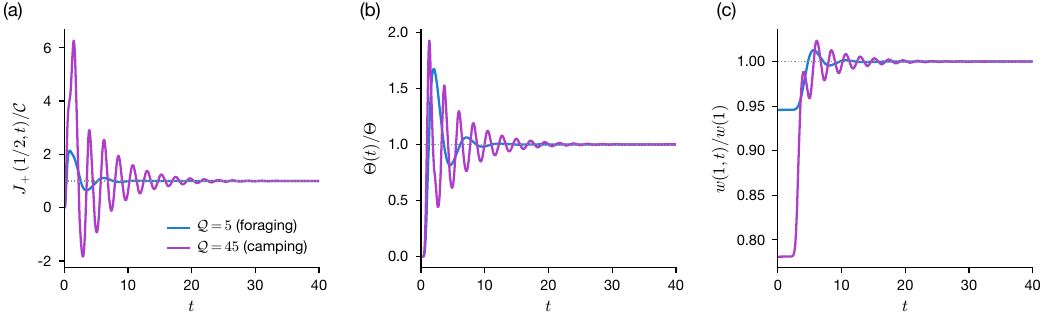}
\caption{Relaxation of (a) the outbound current at mid-trail, $J_+(1/2,t)$, (b) $\Theta(t)$ and (c) $w(1,t)$, each divided by its value in the steady state ($\Cc$, $\Theta$, $w(1)$), at $\varepsilon=0.05$, $g_F=6$, $g_N=2$, $\tilde\kappa=4$, for the foraging ($\Qt=5$) and camping ($\Qt=45$) states. The steady-state values are those of the steady solver, not the end of the run, so convergence to $1$ (dotted) shows that the forward-in-time system relaxes onto the solution of the steady problem. Here $J_+$ is the current of the outbound ant density, $\varrho_+$, evaluated at the midpoint of the trail.}%
\label{fig:dynseries}
\end{figure}

Figure~\ref{fig:dynseries} follows the circulation, $\Theta$ and $w(1)$ in time. The oscillations are complete by $\tilde{t}\simeq20$ in the foraging state and by $\tilde{t}\simeq40$ in the camping state.

\subsection*{Numerical method}

\begin{figure}[H]
\centering
\includegraphics[width=\textwidth]{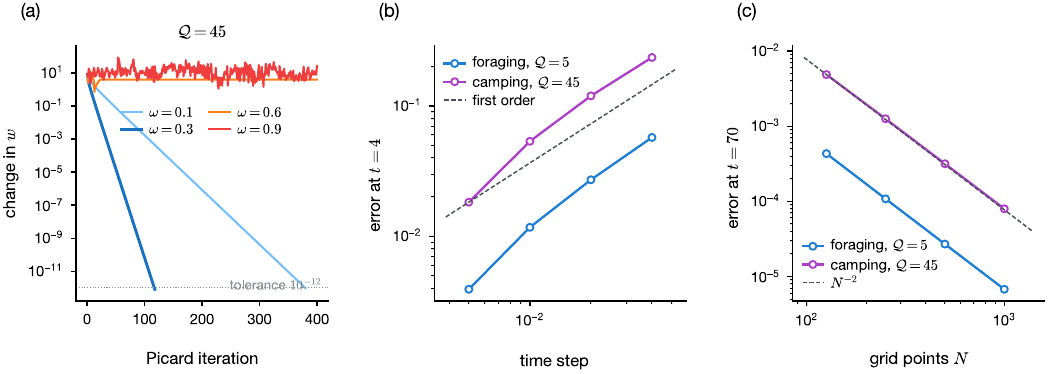}
\caption{Numerics of the one-dimensional trail at $\varepsilon=0.05$,
$g_F=6$, $g_N=2$, $\tilde\kappa=4$. (a)~Steady solver: largest change in $w$ between successive Picard iterations at $\Qt=45$ for four relaxation factors $\omega$; dotted: the required tolerance $10^{-12}$. Forward-in-time solver, for the foraging ($\Qt=5$) and camping ($\Qt=45$) states: (b)~largest difference between the fields $w$, $v$, $\varrho_\pm/\Qt$ and a reference run with a step of $0.0025$, at $t=4$, against the time step (constant step); dashed: first order. (c)~largest difference between the fields $w$, $v$ and their steady-state values at $t=70$, against the number of grid points $N$; dashed: $N^{-2}$.}%
\label{fig:numerics}
\end{figure}

\paragraph*{Steady state.} The system of Eqs.~\eqref{eq:logeqs}--\eqref{eq:logbcs} is solved on a uniform grid of $N$ points and spacing $h$, with the discrete residual
\begin{equation}
R_j=\varepsilon\,\frac{y_{j+1}-2y_j+y_{j-1}}{h^{2}}
+\varepsilon\left(\frac{y_{j+1}-y_{j-1}}{2h}\right)^{2}-y_j+f_j ,
\end{equation}
where $y$ stands for $w$ or $v$ and $f$ for the source of the respective equation. Each field is found by a damped Newton iteration whose Jacobian is tridiagonal, solved by a banded direct solve; Neumann ends are imposed by the ghost-point form $\varepsilon\,2(y_1-y_0)/h^{2}-y_0+f_0$. The coupling is resolved by Picard outer loop:
\begin{enumerate}
\item Solve the $v$ equation, which for the one-way model is independent of everything else;
\item Evaluate $\varrho_-$ from Eq.~\eqref{eq:pminus} and $c$ from Eq.~\eqref{eq:cquad};
\item Solve the $w$ equation with that source;
\item Mix the new and old sources with relaxation factor $\omega=0.3$ and repeat to a residual of $10^{-12}$.
\end{enumerate}
Continuation in $\Qt$ is used throughout: the sweep starts at small $\Qt$ and steps upward, each solve seeded with the previously converged solution. The Newton iteration converges in a few iterations, so the outer loop is the limiting step: at $\Qt=45$ it converges for $\omega=0.1$ and $0.3$, the latter in about $120$ iterations, but does not converge for $\omega=0.6$ and $0.9$ (see Fig.~\ref{fig:numerics}(a)).

\paragraph*{Time dynamics.} The forward-in-time system Eqs.~\eqref{eq:dynfields}--\eqref{eq:dynrho} is advanced by backward Euler. Each step solves the $v$ and $w$ equations by Newton iteration as above, with the time-derivative term added to the diagonal, and the two are coupled through the boundary conditions in Eq.~\eqref{eq:logbcs} by two sweeps per step. The densities are then advanced with the updated fields. The flux between neighboring nodes is the exact flux of Eq.~\eqref{eq:dynrho} for a value field that is linear across the cell, so the zero-current state $\varrho_\pm\propto\chi_\pm^2$ of Eq.~\eqref{eq:boltzmann} is reproduced exactly. Ants arriving at the food are added to $\varrho_-$ and ants arriving at the nest to $\varrho_+$, so the colony size $\Qt$ is conserved by construction. The time step sweep starts at $10^{-3}$ and grows by $5\%$ per step up to a maximum of $0.025$. At fixed $N=501$ the error at $t=4$, measured against a reference run with a step of $0.0025$, falls linearly with the step (see Fig.~\ref{fig:numerics}(b)). In space, the fields at $t=70$ differ from the independently solved steady state by an amount that falls as $N^{-2}$, matching the spatial order of the steady solver (see Fig.~\ref{fig:numerics}(c)).

\section{Trail networks and circuit reduction}%
\label{sm:graph}

\subsection*{Geometry and vertex conditions}

Let $P$ arcs join a nest node $N$ to a food node $F$. Arc $i$ has length $\ell_i$ and is parametrized by arclength $x\in[0,\ell_i]$ with $x=0$ at the nest. Let the reference length be $L=\ell_1$, so that a single non-dimensional diffusion
\begin{equation}
\varepsilon=\frac{D}{\lambda\,\ell_1^{2}}
\end{equation}
serves for the whole network and the scaled arc lengths are
\begin{equation}
L_1=1,\qquad L_i=\ell_i/\ell_1 .
\end{equation}
For the double bridge, $P=2$ and the arcs differ only through $r\equiv\ell_2/\ell_1$; all other parameters $(\varepsilon,g_F,g_N,\tilde\kappa,\Qt)$ are shared, and $r=1$ is the symmetric bridge. The homing potential is a function of the arclength from the nest along the arc, $\Ut_i=\tilde\kappa\,\xt^{2}$.

Continuity of the value fields at each node, together with Kirchhoff matching of the derivatives, gives
\begin{align}
\text{nest }N:&\quad
w_i(0)=w_N\ \ \forall i,\qquad
\sum_i\partial_{\xt} w_i(0)=0,\qquad
v_i(0)=w_N+g_N,
\label{eq:nestvertex}\\[2pt]
\text{food }F:&\quad
v_i(L_i)=v_F\ \ \forall i,\qquad
\sum_i\partial_{\xt} v_i(L_i)=0,\qquad
w_i(L_i)=v_F+g_F .
\label{eq:foodvertex}
\end{align}
Since $Z=e^{w}$ and $\Psi=e^{v}$ with $Z,\Psi>0$, continuity and Kirchhoff matching transfer unchanged between $V^\pm$, $Z,\Psi$ and $w,v$.

Each field satisfies $P$ second-order equations and so needs $2P$ conditions. For $w$: continuity at $N$ supplies $P-1$, Kirchhoff at $N$ supplies $1$, and the Dirichlet boundary condition at each of the $P$ food ends supplies $P$, giving $2P$. For $v$ the mirror count applies at $F$. The mirror structure of the single trail therefore survives: $w$ carries Kirchhoff at the nest and clamps at the food, and $v$ the reverse.

\subsection*{Four currents}

Away from the two conversion nodes there are no sources, so in steady state $\partial_{\xt} J_{\pm,i}=0$ on each arc and the currents are constant along it,
\begin{equation}
J_{+,i}=+\Cc_i^{+},\qquad J_{-,i}=-\Cc_i^{-} .
\end{equation}
In the scaled logarithmic variables the outbound current relation Eq.~\eqref{eq:cdef} reads on arc $i$
\begin{equation}
\varepsilon\Big(2\varrho_{+,i}\,w_i' - \varrho_{+,i}'\Big)=\Cc_i^{+} .
\label{eq:arccurrent}
\end{equation}
Repeating the calculation of Eq.~\eqref{eq:ohmscaled} arc by arc, with $\varphi_{+,i}=\varepsilon\varrho_{+,i}e^{-2w_i}$, gives Ohm's law in differential form,
\begin{equation}
-\,\partial_{\xt}\varphi_{+,i}\;=\;\Cc_i^{+}\,e^{-2w_i} ,
\qquad
+\,\partial_{\xt}\varphi_{-,i}\;=\;\Cc_i^{-}\,e^{-2v_i} .
\label{eq:ohmarc}
\end{equation}
Here $\varphi_\pm$ is the potential, which drops in the direction of current flow; $\Cc_i^\pm$ is the current, constant along the arc; and $e^{-2w}=Z^{-2}$ is the local resistivity.

\paragraph{Physical meaning of the potential.} From Sec.~\ref{sm:fpe}, $\varrho_+\propto Z^{2}$ is the zero-current Gibbs state, so $\varphi_+$ measures the density relative to its local equilibrium weight. It is the excess over detailed balance, and it is this excess that drives current. Where the solution is locally Gibbs, $\varphi_+$ is constant and nothing flows. The analogy is with a chemical potential difference driving a flux.

\paragraph{Trail as a conductor.} By Sec.~\ref{sm:positivity} the resistivity satisfies $0<e^{-2w}\le1$, and it is exponentially small wherever the trail is strong. A well-marked trail is an exponentially good conductor, and this exponential sensitivity is the amplification mechanism of the model.

\subsection*{Arc resistance}

Integrating Eq.~\eqref{eq:ohmarc} between two points $\xt<\xt'$ on the same arc,
\begin{equation}
\varphi_{+}(\xt)-\varphi_{+}(\xt')=\Cc^{+}\!\int_{\xt}^{\xt'}\!e^{-2w}\,\dd\tilde y ,
\end{equation}
which is the finite form of Ohm's law with resistance $\int_{\xt}^{\xt'}e^{-2w}$. At the food end $\varrho_{+}(L_i)=0$, so $\varphi_{+}(L_i)=0$ and the food node is the ground. Hence $\varphi_{+,i}(0)=\Cc_i^{+}R^+_i$ with the arc resistances
\begin{equation}
R^+_i\equiv\int_{0}^{L_i}e^{-2w_i}\,\dd\xt, \qquad
R^-_i\equiv\int_{0}^{L_i}e^{-2v_i}\,\dd\xt .
\label{eq:resistdef}
\end{equation}
Restoring the density,
\begin{equation}
\varrho_{+,i}(\xt)=\frac{\Cc_i^{+}}{\varepsilon}\,e^{2w_i(\xt)}
\!\int_{\xt}^{L_i}\!e^{-2w_i}\,\dd\tilde y,
\qquad
\varrho_{+,i}(0)=\frac{\Cc_i^{+}}{\varepsilon}\,e^{2w_N}R^+_i ,
\label{eq:rhoplusarc}
\end{equation}
using $w_i(0)=w_N$ from Eq.~\eqref{eq:nestvertex}. The returner calculation is identical with $v$ in place of $w$ and the nest as ground,
\begin{equation}
\varrho_{-,i}(\xt)=\frac{\Cc_i^{-}}{\varepsilon}\,e^{2v_i(\xt)}
\!\int_0^{\xt}\!e^{-2v_i}\,\dd\tilde y,
\qquad
\varrho_{-,i}(L_i)=\frac{\Cc_i^{-}}{\varepsilon}\,e^{2v_F}R^-_i .
\label{eq:rhominusarc}
\end{equation}

\subsection*{Splitting law}

We can impose continuity of $\varrho_+$ at the nest by using Eq.~\eqref{eq:rhoplusarc} for the common value $\varrho_+(0)=\dfrac{\Cc_i^{+}}{\varepsilon}e^{2w_N}R^+_i$ for every $i$. So the arc dependence vanishes and we get:
\begin{equation}
\Cc_1^{+}R^+_1=\Cc_2^{+}R^+_2=\cdots=\varepsilon\,\varrho_{+}(0)\,e^{-2w_N}
\;\equiv\;\varphi_{+}(0).
\end{equation}
The factors $\varepsilon$ and $e^{2w_N}$ drop out because they are properties of the node, not of the arc. Writing this as $\Cc_i^{+}=\varphi_{+}(0)/R^+_i$ is Ohm's law for arcs in parallel across a common potential drop, with $\varphi_+(0)$ the electromotive force supplied by the nest. Applying the same argument to $\varrho_-$ at the food gives $\Cc_1^{-}R^-_1=\Cc_2^{-}R^-_2=\dots$, and imposing conservation at the two conversion nodes, $\sum_i \Cc_i^{+}=\sum_i \Cc_i^{-}=\ctot$, fixes the normalization,
\begin{equation}
\Cc_i^{+}=\ctot\,\frac{(R^+_i)^{-1}}{\sum_j (R^+_j)^{-1}},
\qquad
\Cc_i^{-}=\ctot\,\frac{(R^-_i)^{-1}}{\sum_j (R^-_j)^{-1}} ,
\label{eq:ohmSM}
\end{equation}
which is Eq.~\eqref{eq:ohm} of the article. These are two independent parallel-resistor networks sharing a total current: outbound ants divide by $Z$-resistance and returners by $\Psi$-resistance. Nothing in the derivation used $P=2$, so Eq.~\eqref{eq:ohmSM} holds for any number of arcs. The total current follows from the population constraint,
\begin{equation}
\sum_i\int_{0}^{L_i}\big(\varrho_{+,i}+\varrho_{-,i}\big)\,\dd\xt=\Qt .
\end{equation}

\paragraph{Need for four currents.} Suppose we had assumed a single current per arc, $J_+=+\Cc_i$ and $J_-=-\Cc_i$, as holds on a single trail. Continuity of $\varrho_+$ at the nest would give $\Cc_1 R^+_1 = \Cc_2 R^+_2$ and continuity of $\varrho_-$ at the food would give $\Cc_1 R^-_1 = \Cc_2 R^-_2$. These are two conditions on the single ratio $\Cc_1/\Cc_2$, compatible only in the non-generic case $R^+_1/R^-_1=R^+_2/R^-_2$, so the single-current network is over-determined. The physical content of the failure is that an ant which reaches the food along arc~1 is under no obligation to return along arc~1. Outbound and returning traffic split independently, and Eq.~\eqref{eq:ohmSM} is the statement that each split is set by its own resistance network. On each arc $J_++J_-=\Cc_i^{+}-\Cc_i^{-}$, which no longer vanishes pointwise, although $\sum_i(\Cc_i^{+}-\Cc_i^{-})=0$. This is a genuine circulation around the loop and is the signature of a circuit rather than of a dead-ended segment.

\paragraph{Series and parallel.} Splitting an arc at an interior point $\xt_*$ gives $R=\int_0^{\xt_*}e^{-2w}+\int_{\xt_*}^{L_i}e^{-2w}$ with the same current through both segments, so the segments behave as resistors in series. Eq.~\eqref{eq:ohmSM} is the parallel rule, with effective bundle resistance $R_{\rm eff}^{-1}=\sum_j R_j^{-1}$. Together these mean that an arbitrary network, not merely the double bridge, reduces to elementary circuit algebra once the fields $w,v$ are known, with no further partial differential equations to solve. The whole nonlinearity of the problem sits in the fact that the resistivity $e^{-2w}$ depends on the current the arc is carrying.

\section{Why one-way marking cannot break symmetry}%
\label{sm:nullresult}

Before extending the model with two-way marking of pheromones, it is worth establishing what a one-way marking does not contain. We can trace the dependencies through the circuit:
\begin{equation}
\Ut_i\ \longrightarrow\ v_i\ \longrightarrow\ R^-_i\ \longrightarrow\
\Cc_i^{-}\ \longrightarrow\ \varrho_{-,i}\ \longrightarrow\ w_i\
\longrightarrow\ R^+_i\ \longrightarrow\ \Cc_i^{+}.
\label{eq:openchainSM}
\end{equation}
The chain is read as follows: the $v$ equation of Eq.~\eqref{eq:logeqs} contains no density at all, so $v_i$ is determined by $\Ut_i$ and the arc length alone. This fixes $R^-_i$ by Eq.~\eqref{eq:resistdef}, hence $\Cc_i^{-}$ by Eq.~\eqref{eq:ohmSM}, hence $\varrho_{-,i}$ by Eq.~\eqref{eq:rhominusarc}. The returner density then sources the $w$ equation, which fixes $R^+_i$ and finally $\Cc_i^{+}$.

Two consequences follow:
\paragraph{At $r=1$ the symmetric state is the unique state, at every $\Qt$.} The $v$ problem has data symmetric under arc exchange, so by uniqueness of its boundary-value problem $v_1\equiv v_2$. Hence $R^-_1=R^-_2$, hence $\varrho_{-,1}\equiv\varrho_{-,2}$, hence $w_1\equiv w_2$, hence $R^+_1=R^+_2$, and both currents split evenly. No selection is possible at any colony size.

\paragraph{At $r\neq1$ the split is fixed by resistance.} Since $\Cc_i^{-}\propto (R^-_i)^{-1}$ is fixed by geometry once and for all, the split cannot depend on history, so there is no bistability at any asymmetry. The structural reason is that in the one-way model only outbound ants sense the trail and only returners deposit it. Autocatalysis requires a closed loop, and Eq.~\eqref{eq:openchainSM} is as of now open.

\section{Bidirectional marking}%
\label{sm:bidirectional}

\subsection*{Biological input}

Goss \emph{et al.} report that \emph{L.\ humile} workers mark both when leaving and when returning to the nest, during food and exploratory recruitment, and that their choice function is applied at both choice points to ants travelling in either direction~\cite{Goss1989}. This species both deposits and reads in both directions. There is one chemical field, and every ant contributes to it and responds to it. This statement introduces exactly two parameters.

\subsection*{Mark field and running costs}

Let returners deposit at unit rate and outbound ants at relative rate $\sigma\ge0$, so that the local density of marks is
\begin{equation}
m(x)\;=\;\varrho_-(x)+\sigma\,\varrho_+(x).
\label{eq:markfield}
\end{equation}
Both species are rewarded for being where marks are, outbound ants with weight $\alpha$ and returners with weight $\alpha_-$, so the running costs of Sec.~\ref{sm:hjb} become
\begin{equation}
f_+=-\alpha\,m,
\qquad
f_-=U-\alpha_-\,m .
\label{eq:runningcosts}
\end{equation}
The one-way model of the first part of the article is $\sigma=0$, $\alpha_-=0$.

\subsection*{Modified field equations}

The HJB equations pick up the new terms directly:
\begin{align}
\partial_t V^{+}+D\nabla^2V^{+}-\tfrac{\beta}{2}|\nabla V^{+}|^{2}-\lambda V^{+}
-\alpha m+\Gamma_+\big(V^{-}-V^{+}-r_F\big)&=0,\\
\partial_t V^{-}+D\nabla^2V^{-}-\tfrac{\beta}{2}|\nabla V^{-}|^{2}-\lambda V^{-}
+U-\alpha_- m+\Gamma_-\big(V^{+}-V^{-}-r_N\big)&=0 .
\end{align}
Applying the Cole--Hopf map $V^{+}=-\dfrac{2D}{\beta}\ln Z$, $V^{-}=-\dfrac{2D}{\beta}\ln\Psi$ and multiplying through by $-\beta Z/2D$ and $-\beta\Psi/2D$ respectively, exactly as in Sec.~\ref{sm:colehopf}, the steady instantaneous-switching equations are
\begin{align}
D\nabla^2Z-\lambda Z\ln Z+\frac{\alpha\beta}{2D}\,m\,Z&=0,\\
D\nabla^2\Psi-\lambda\Psi\ln\Psi-\frac{\beta}{2D}U\Psi
+\frac{\alpha_-\beta}{2D}\,m\,\Psi&=0 .
\end{align}
Non-dimensionalising as in Sec.~\ref{sm:onedim}, the scaling that defines $\varrho_\pm$ gives
\begin{equation}
\frac{\alpha\beta}{2D\lambda}\,m
=\frac{\alpha\beta}{2D\lambda}\big(\varrho_-+\sigma\varrho_+\big)
=\varrho_-+\sigma\varrho_+\;\equiv\;M ,
\end{equation}
and, writing $\sigma_-\equiv\alpha_-/\alpha$,
\begin{equation}
\frac{\alpha_-\beta}{2D\lambda}\,m
=\frac{\alpha_-}{\alpha}\cdot\frac{\alpha\beta}{2D\lambda}\,m
=\sigma_-\,M .
\end{equation}
Hence the extended dynamics
\begin{equation}
\begin{aligned}
\varepsilon\big(w''+w'^{2}\big)-w+M&=0,\\[2pt]
\varepsilon\big(v''+v'^{2}\big)-v-\Ut+\sigma_-M&=0,
\end{aligned}
\qquad
M\equiv\varrho_-+\sigma\,\varrho_+ .
\label{eq:bidirfieldsSM}
\end{equation}

\paragraph{Additional parameters.} Both are assumed to be of order unity for \emph{L.\ humile}. The quantity $\sigma$ is the deposition asymmetry, measuring how much outbound ants mark relative to returners, and $\sigma_-$ is the sensing asymmetry, measuring how strongly returners respond to marks relative to outbound ants. Setting either to zero re-opens the feedback chain Eq.~\eqref{eq:openchainSM}, so both need to be retained.

\paragraph{Deposition--sensing matrix.} Substituting $M$ into the second line of Eq.~\eqref{eq:bidirfieldsSM} leaves the compound coefficient $\sigma_-\sigma$ multiplying $\varrho_+$. Naming it
\begin{equation}
\eta\;\equiv\;\sigma_-\,\sigma
\label{eq:etadef}
\end{equation}
gives the two field equations with one explicit coefficient per density,
\begin{equation}
\begin{aligned}
\varepsilon\big(w''+w'^{2}\big)-w
   \;+\;\varrho_-\;+\;\sigma\,\varrho_+ &=0,\\[2pt]
\varepsilon\big(v''+v'^{2}\big)-v-\Ut
   \;+\;\sigma_-\,\varrho_-\;+\;\eta\,\varrho_+ &=0 .
\end{aligned}
\label{eq:bidirfields2}
\end{equation}
The four coefficients $(1,\sigma;\sigma_-,\eta)$ form the deposition--sensing matrix: row one is what outbound ants read and row two what returners read, while column one is what returners deposit and column two what outbound ants deposit. The cross-coupling $\eta$ is the term that closes the feedback loop through both channels at once, and it is the natural object to quote when discussing the instability.

Note that $\eta$ is not an independent parameter. It is fixed by $\sigma$ and $\sigma_-$ through Eq.~\eqref{eq:etadef}, and the continuation code carries $\sigma$ and $\sigma_-$ as the two independent inputs. Eq.~\eqref{eq:bidirfields2} is a rewriting for readability and not a generalization; treating $\eta$ as free would give a strictly larger model in which deposition and sensing no longer factorize.

\paragraph{Full system.} The transport equations Eq.~\eqref{eq:currents} and the constancy of $J_\pm$ along an arc are same as in Eq.~\eqref{eq:runningcosts}, so the entire circuit reduction of Sec.~\ref{sm:graph} carries over verbatim. Only the field equations that determine $w$ and $v$, and hence the resistances $R^+_i$ and $R^-_i$, are modified. What changes is that $R^-_i$ now depends on the densities and therefore on the currents, which is exactly the feedback path missing from Eq.~\eqref{eq:openchainSM}.

With $\sigma_->0$ the returner equation acquires a positive source, so $v$ is no longer negative throughout the bulk and $\Psi$ may exceed unity away from the nest. The positivity argument of Sec.~\ref{sm:positivity} applies to $Z$ and is unaffected, since the mark field $m\ge0$ enters the $Z$ equation with the same sign as $\varrho_-$.

\section{Numerical continuation on the double bridge}%
\label{sm:auto}

The bidirectional model in Eq.~\eqref{eq:bidirfieldsSM} is solved by pseudo-arclength continuation in AUTO-07p~\cite{Auto07p}. This section details the formulation, implementation and the findings.

\subsection*{Formulation}

The first step in the continuation is to define the potentials of Sec.~\ref{sm:graph} as state variables. Ohm's law in differential form,
\begin{equation}
\frac{\dd\varphi_+}{\dd \xt}=-\Cc^{+}e^{-2w},
\qquad
\frac{\dd\varphi_-}{\dd \xt}=+\Cc^{-}e^{-2v},
\end{equation}
turns the closed-form density integrals in Eqs.~\eqref{eq:pplus}--\eqref{eq:pminus} into local ordinary differential equations, so the whole problem becomes a two-point boundary-value problem with no integrals in the right-hand side. The densities are recovered pointwise as $\varrho_+=\varphi_+e^{2w}/\varepsilon$ and $\varrho_-=\varphi_-e^{2v}/\varepsilon$.

Per arc, on $t\in[0,1]$ with $\xt=L_i t$, the state is
\begin{equation}
\big(w,\;p=w',\;v,\;q=v',\;\varphi_+,\;\varphi_-,\;\xt,\;Q,\;A,\;B\big),
\end{equation}
ten components, so $\mathrm{NDIM}=20$ for two arcs. The arclength $\xt$ is carried explicitly as a state because AUTO's right-hand side is not given the independent variable and the homing potential $\Ut=\tilde\kappa\,\xt^{2}$ needs it. The remaining three are accumulators, $Q$ for the population and $A,B$ for the two resistances, each satisfying $\dd Q/\dd t=L_i(\varrho_++\varrho_-)$, $\dd A/\dd t=L_i e^{-2w}$ and $\dd B/\dd t=L_i e^{-2v}$ with zero initial value, so that their terminal values are the integrals in Eq.~\eqref{eq:resistdef}.

There are $\mathrm{NBC}=26$ boundary conditions. Ten per arc:
\begin{equation}
\begin{aligned}
&w_i(0)=w_N, && v_i(0)=w_N+g_N, && \varphi_{-,i}(0)=0,\\
&\xt_i(0)=0, && Q_i(0)=R^+_i(0)=R^-_i(0)=0, &&\\
&w_i(L_i)=v_F+g_F, && v_i(L_i)=v_F, && \varphi_{+,i}(L_i)=0,
\end{aligned}
\end{equation}
which are the continuity and Dirichlet conditions in Eqs.~\eqref{eq:nestvertex}--\eqref{eq:foodvertex}, the two grounds, and the four accumulator initialisations. Six coupling conditions close the system:
\begin{equation}
\begin{aligned}
&\textstyle\sum_i\partial_{\xt} w_i(0)=0, &&\textstyle\sum_i\partial_{\xt} v_i(L_i)=0
&&\text{(Kirchhoff at the two nodes)},\\
&\varphi_{+,1}(0)=\varphi_{+,2}(0), &&\varphi_{-,1}(L_1)=\varphi_{-,2}(L_2)
&&\text{(common node potentials)},\\
&\Cc_1^{+}+\Cc_2^{+}=\Cc_1^{-}+\Cc_2^{-}, && Q_1(L_1)+Q_2(L_2)=\Qt
&&\text{(conversion balance, population)}.
\end{aligned}
\end{equation}
This requires $\mathrm{NBC}-\mathrm{NDIM}+1=7$ free parameters: the continuation parameter together with $w_N$, $v_F$, $\Cc_1^{\pm}$ and $\Cc_2^{\pm}$. That the count balances exactly is a useful check on the vertex conditions.

\paragraph{Ohm's law.} Because $\varphi_{+,i}(0)=\Cc_i^{+}R^+_i$ follows from integrating the potential equation, the boundary condition that the two arcs share a nest potential, $\varphi_{+,1}(0)=\varphi_{+,2}(0)$, \emph{is} the splitting law $\Cc_1^{+}R^+_1=\Cc_2^{+}R^+_2$. The same holds at the food for the returners. The circuit laws of Sec.~\ref{sm:graph} therefore enter the computation as boundary conditions on the potential rather than as separate algebraic constraints.

\paragraph{Initial solution.} At $\Qt=g_F=g_N=\tilde\kappa=0$ the exact solution is
\begin{equation}
w\equiv v\equiv 0,\qquad \varphi_\pm\equiv0,\qquad
\Cc_i^{\pm}=0,\qquad w_N=v_F=0,
\end{equation}
which satisfies the field equations, both clamps and Kirchhoff identically: $w\equiv0$ solves $\varepsilon(w''+w'^2)-w+M=0$ with $M=0$, and meets $w(0)=w_N=0$, $w(1)=v_F+g_F=0$ and $p(0)=0$. The continuation therefore starts already converged, and the physical operating point is reached by switching the parameters on one at a time.

\paragraph{Operating point and findings.} The ladder raises $\Qt$, then $g_F=6$, $\tilde\kappa=4$, $g_N=2$ and $\sigma_-=1$, at $\varepsilon=0.05$. Figure~\ref{fig:bridge} uses $\Qt=2.154$, chosen so that the broken branch peaks at $|\asym|=0.60$, i.e.\ $\traf=80\%$. At $r=1$ the symmetric branch folds at $\sigma=0.876$ and $0.907$ and has its branch point at $\sigma_c=0.8954$ between them. The broken branch leaves $\sigma_c$ towards smaller $\sigma$, turns at a fold at $\sigma=0.8905$ ($|\asym|=0.48$) and reaches $|\asym|=0.600$ at $\sigma\simeq1.00$; near the branch point $\sigma-\sigma_c\approx-0.0070\,\asym^2$, and the local exponent of $|\asym|$ against $|\sigma-\sigma_c|$ is $0.496$ for $|\asym|<0.05$. A scan of $\Qt$ shows the peak rising steadily with colony size, from $74.7\%$ at $\Qt=2$, where the pitchfork is supercritical, to $92\%$ at $\Qt=3$; the pitchfork is subcritical at $\Qt=2.04$.

\section{Comparison with the double-bridge experiments}%
\label{sm:goss}

\subsection*{Experiments}

Goss \emph{et al.}~\cite{Goss1989} performed experiments in laboratory colonies of \emph{Iridomyrmex}/\emph{Linepithema humilis} that can access food through a bridge of two identical shapes, each offering a short and a long branch of length ratio $r$. Branches meet the bridge axis at $30^\circ$ so that the forager has no positional preference, and one module's short branch is mirrored relative to the other's to cancel any external bias. Selection becomes visible four to eight minutes after the bridge is placed. Per-trial traffic is reported as between $317$ and $1043$ ants over ten minutes, that is $0.5$--$1.7\,\mathrm{s}^{-1}$, and selection reliability increases with traffic.

\subsection*{Primary observations}

\begin{enumerate}[leftmargin=2em,itemsep=3pt]
\item \textbf{$r=1$: no preference.} $12/26$ trials, $p>0.05$. Each trial selects a branch strongly, but which one is random across trials.
\item \textbf{$r=1.4$: significant selection.} $15/18$ choose the short branch, $p<0.05$.
\item \textbf{$r=2.0$: complete selection.} $14/14$, $p<0.05$.
\item \textbf{$r=2.0$ with the short branch added late: trapping.} When the short branch is introduced $30$~min after the trial begins and traffic is measured $20$--$30$~min later, only $2/18$ colonies switch to it, $p<0.05$. The other sixteen stay on the long branch.
\item \textbf{Selection strength.} In $11$ of $14$ experiments at $r=2$, more than $80\%$ of total traffic used the short branch.
\end{enumerate}

\subsection*{Proposed mechanism by Goss et al.}

Our theory and the model in Ref.~\cite{Goss1989} agree on the sign of the effect but phenomenological assumptions are different. Their model is,
\begin{equation}
\frac{\dd S_j}{\dd t}=\Phi p_{s,j'}(t-20)+\Phi p_{s,j}(t),
\qquad
\frac{\dd L_j}{\dd t}=\Phi p_{l,j'}(t-20r)+\Phi p_{l,j}(t),
\label{eq:gossmodel}
\end{equation}
where $S_j$ and $L_j$ with choice function
\begin{equation}
p_{s,j}=\frac{(20+S_j)^{2}}{(20+S_j)^{2}+(20+L_j)^{2}}, 
\label{eq:gosschoice}
\end{equation}
selects by transit delay: an ant crossing the short branch reaches the far choice point after $20$~s and one crossing the long branch after $20r$~s, so in the interval between the short branch is marked from both ends while the long branch is marked from one. Our length dependence enters instead through the stationary resistance $R^+_i=\int e^{-2w_i}\sim L_i e^{-2\langle w_i\rangle}$ of Sec.~\ref{sm:graph}, with no reference to transit time.

Eq.~\eqref{eq:gossmodel} also does not have a decay term, evaporation being dropped on the grounds that the pheromone lifetime of about $30$~min exceeds the experiment. Consequently $S_j$ and $L_j$ grow without bound and the model has no stationary state: selection there is a purely transient ratchet and trapping is permanent by construction. The present theory is a steady-state theory with evaporation built in, so trapping has to be earned and it persists only inside the bistable region and its residence time is finite.

\end{document}